\documentclass[journal, letterpaper]{IEEEtran}
\usepackage[dvips]{graphicx}
\usepackage[cmex10]{amsmath}
\usepackage{multirow}
\usepackage{epsfig}
\usepackage{mathrsfs}
\usepackage{amssymb}
\usepackage{comment}
\usepackage{bm}
\usepackage{dsfont}
\usepackage{array}
\usepackage{amsthm}
\usepackage{blkarray}
\usepackage{enumerate}
\usepackage{url}
\usepackage{chemarrow}
\usepackage{color}
\usepackage[lined,ruled,linesnumbered,noend]{algorithm2e}
\usepackage{epsf,psfrag}
\usepackage{epsfig}

\def\Xc{\mathcal{X}}
\def\Lc{\mathcal{L}}
\def\Gc{\mathcal{G}}
\def\xbf{\mathbf{x}}

\def\uML{u^{\mbox{\tiny ML}}}

\def\argmax{\mathop{\arg\max}}
\def\argmin{\mathop{\arg\min}}
\def\tmix{t_{\mbox{\tiny mix}}}
\def\mbbE{\mathbb{E}}
\def\cmin{c_{\min}}
\def\cmax{c_{\max}}
\def\pimax{\pi_{\max}}

\def\pmax{p_{\max}}
\def\pmin{p_{\min}}
\def\PIM{P_{\mbox{\tiny IM}}} 
\def\PML{P_{\mbox{\tiny ML}}} 
\def\POO{P_{\mbox{\tiny OO}}} 
\def\PMO{P_{\mbox{\tiny MO}}} 
\def\PCML{P_{\mbox{\tiny CML}}} 
\def\psiMY{\psi^{\mbox{\tiny MY}}} 

\theoremstyle{definition}
\newtheorem{theorem}{Theorem}[section]
\newtheorem{lemma}[theorem]{Lemma}
\newtheorem{corollary}[theorem]{Corollary}

\ifCLASSINFOpdf

\else

\fi

\usepackage[tight,footnotesize]{subfigure}

\begin{document}

\IEEEoverridecommandlockouts
%
\title{Location Privacy in Mobile Edge Clouds: A Chaff-based Approach}
\author{
Ting He, Ertugrul N. Ciftcioglu, Shiqiang Wang, and Kevin S. Chan
\thanks{
T. He is with Pennsylvania State University, University Park, PA, USA. Email: tzh58@psu.edu

E. N. Ciftcioglu and K. S. Chan are with Army Research Laboratory, Adelphi, MD, USA. Email: ertugrulnc@ieee.org, kevin.s.chan.civ@mail.mil

S. Wang is with IBM T. J. Watson Research Center, Yorktown, NY, USA. Email: wangshiq@us.ibm.com

 Preliminary results of this work have been presented at ICDCS'17 \cite{He17ICDCS}. This paper presents a more comprehensive discussion beyond the conference version, including online user strategies, robustness analysis and defense against advanced eavesdroppers, and performance evaluation driven by real-world user traces.}
}

\maketitle

\begin{abstract}
In this paper, we consider user location privacy in mobile edge clouds (MECs). MECs are small clouds deployed at the network edge to offer cloud services close to mobile users, and many solutions have been proposed to maximize service locality by migrating services to follow their users. Co-location of a user and his service, however, implies that a cyber eavesdropper observing service migrations between MECs can localize the user up to one MEC coverage area, which can be fairly small (e.g., a femtocell). We consider using chaff services to defend against such an eavesdropper, with focus on strategies to control the chaffs. Assuming the eavesdropper performs maximum likelihood (ML) detection, we consider both heuristic strategies that mimic the user's mobility and optimized strategies designed to minimize the detection or tracking accuracy. We show that a single chaff controlled by the optimal strategy or its online variation can drive the eavesdropper's tracking accuracy to zero when the user's mobility is sufficiently random. We further propose extended strategies that utilize randomization to defend against an advanced eavesdropper aware of the strategy. The efficacy of our solutions is verified through both synthetic and trace-driven simulations. \looseness=-1
\end{abstract}

\IEEEpeerreviewmaketitle

\section{Introduction}

While improvement in the coverage of wireless communications brings tons of useful applications to the fingertips of mobile users, this trend also imposes a significant threat on user location privacy. \emph{Location privacy} refers to safeguarding a mobile user's location from unintended use. While legitimate use of user location can enable various \emph{location-based services (LBS)}, malicious use of this information can cause harmful consequences such as stalking, blackmailing, and fraud \cite{Gruteser03MobiSys}.

Existing efforts in protecting user location privacy mostly focus on protecting the information released through the \emph{direct channel}, i.e., location information intentionally revealed by the user. Since the direct channel is controlled by the user, e.g., by configuring whether/when to share his location with an LBS provider, the user can easily obfuscate his location in the spatial/temporal domain to make sure that his location cannot be distinguished from the locations of many other users \cite{Gruteser03MobiSys}.

The more challenging problem, however, is how to prevent unintentional release of location information through \emph{side channels}. In wireless networks, an important side channel is the user's wireless transmission activity, which can be monitored by an eavesdropper with wireless sensing capabilities to track the user, even if the direct channel is perfectly protected. Since its discovery, a few solutions have been proposed to protect this side channel, e.g., by introducing intermittent radio silence and reducing the transmission power, which can effectively increase the uncertainty for the eavesdropper \cite{Hu05SIGCOMM,Jiang07MobiSys}. \looseness=-1

\begin{figure}[tb]
\vspace{-.5em}
\centering
\includegraphics[width=.75\linewidth]{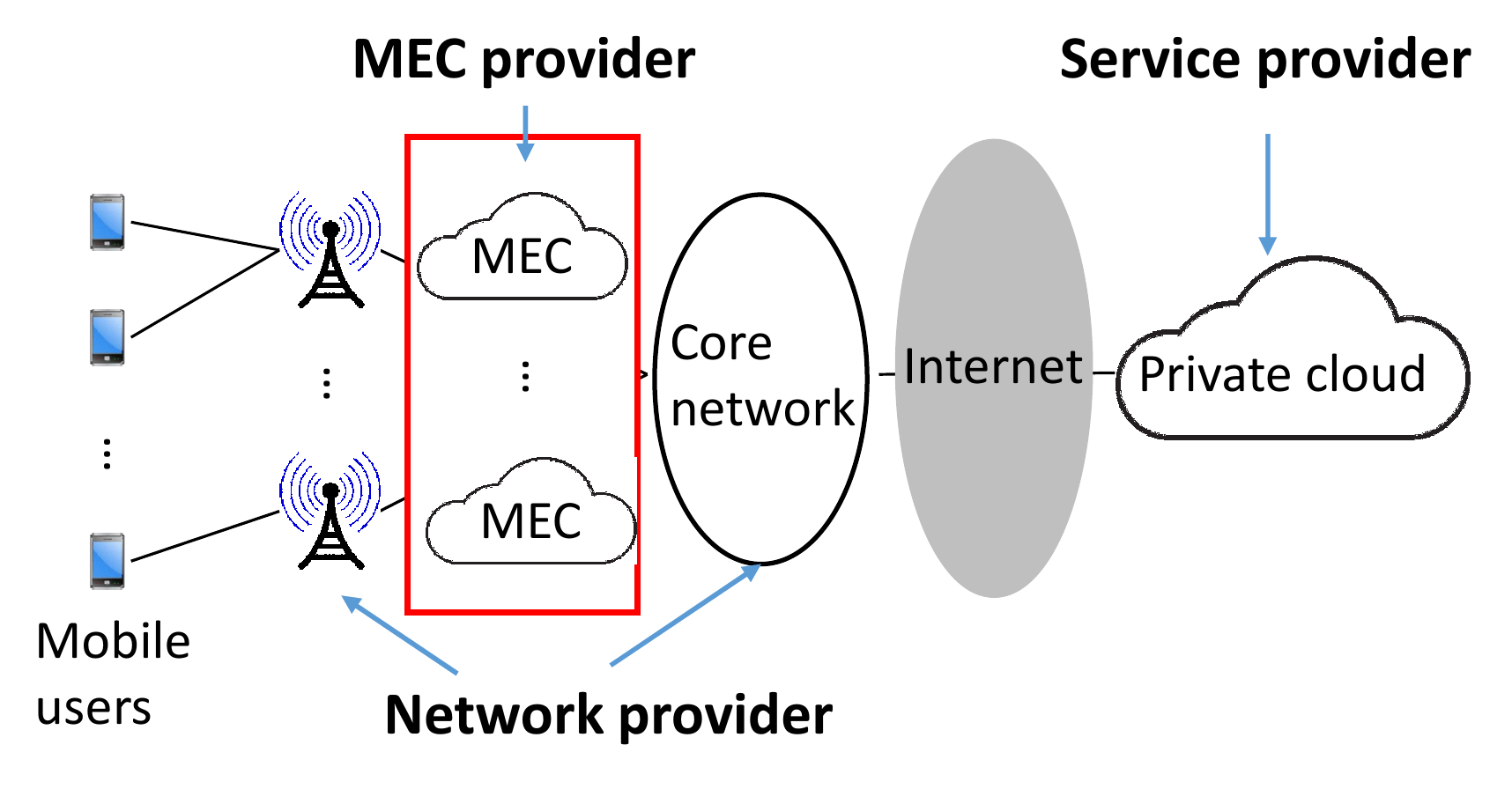}
\vspace{-1em}
\caption{Providing services to mobile users via MECs. } \label{fig:MEC_architecture}
\vspace{-.5em}
\end{figure}

In this work, we investigate the problem in a novel context of \emph{mobile edge clouds (MECs)} \cite{Wang15Networking}, also known as \emph{cloudlets} \cite{Satyanarayanan13PerCom}, \emph{mobile micro clouds} \cite{Wang16TPDS}, fog \cite{Bonomi12MCC}, or \emph{follow me cloud} \cite{Taleb13Network}. As illustrated in Fig.~\ref{fig:MEC_architecture}, MECs are small clouds that offer a selected set of cloud services from the edge of the mobile network (e.g., base stations). Since its introduction, MEC has attracted tremendous interest from both research communities and industry leaders as a promising approach to improve the performance of cloud services for mobile users  \cite{Satyanarayanan15CommMag,IBM13WhitePaper}. It is also considered the most viable approach to offer cloud services in tactical environments \cite{DRDCreport}. From the perspective of location privacy, however, this technology opens a new side channel, referred to as the \emph{cyber side channel}.

Specifically, to deliver the promised performance, MECs need to migrate services\footnote{Here we use ``service'' to refer to an instance of a given type of service (e.g., a VM instance running the service), which is independently generated/migrated for each user of this type of service as in existing solutions \cite{Ha15TechReport,Taleb13Network,Taleb16TCC,Wang15Networking,Wang16TPDS}. } (e.g., by migrating virtual machines (VMs) encapsulating the services \cite{Ha15TechReport}) to follow the mobile users \cite{Taleb13Network,Taleb16TCC,Wang15Networking,Wang16TPDS}. For delay-sensitive services (e.g., augmented reality), the service has to remain no more than one hop away from its user to prevent significant degradation in the Quality of Service (QoS) \cite{Ha15TechReport}.
This implies that a ``cyber eavesdropper'', who can observe service migrations among MECs, can track physical movements of the user. Such a cyber eavesdropper can be a hacker that has gained access to the MECs, or an untrusted MEC provider interested in tracking users of certain services. For example, as tactical operations start to use commercial clouds to reduce cost \cite{DoDCloudStrategy}, a malicious (or compromised) MEC provider can track tactical users by tracking their services.
Cyber eavesdropping is a realistic concern in MECs because of the openness of the MEC ecosystem \cite{Satyanarayanan15CommMag,Schuster16}, which increases the risk of introducing both unsecured systems vulnerable to attacks and untrusted providers.
Note that we distinguish between the network provider, the MEC provider, and the service provider, where the network provider and the service provider are considered as secure and trusted, but the MEC provider can be insecure and untrusted due to the openness of the MEC ecosystem.
The role of the MEC provider is to provide the MEC platform that runs services provided by the service provider, where the MECs and users are interconnected by the network provided by the network provider.
Although the spatial resolution of cyber eavesdropping is limited to the coverage of one MEC (e.g., a cell sector), its harm can be severe, as it can be performed without any physical sensing devices, thus potentially at a much lower cost and a much larger scale.


While cyber eavesdropping and wireless eavesdropping are conceptually similar, the defense mechanisms are quite different. Specifically, the existing defense mechanisms for wireless eavesdropping \cite{Hu05SIGCOMM,Jiang07MobiSys} are \emph{intrusive} in that they modify the user's transmissions. While it is possible to defend against cyber eavesdropping by stopping the service from following the user, such a mechanism will significantly degrade the QoS for delay-sensitive services \cite{Ha15TechReport}. Instead, we consider a \emph{non-intrusive} mechanism using \emph{chaffs}.
Chaffs are legitimate services launched by the user (or by the network provider on behalf of the user) together with the real service to confuse the eavesdropper about which service the user is actually using. For example, they can be implemented by sending fake service requests and handoff signals to user-specified MECs; see Section~\ref{subsec:Eavesdropper Model} for details.

To confuse the eavesdropper, the chaffs must be indistinguishable from the real service. In terms of content, this can be achieved by using independent instances of the same type of service as chaffs.
It is, however, insufficient to only make the chaffs indistinguishable in content. For example, a chaff that never migrates can be easily distinguished from a real service that migrates with the mobile user, and a chaff that randomly migrates among MECs can be easily distinguished from a real service that exhibits temporal correlation in its locations. For a chaff to effectively confuse the eavesdropper, its mobility pattern, i.e., where it is launched and whether/where it is migrated, has to resemble the mobility pattern of the real service. Meanwhile, a chaff that always follows the real service (which follows the user) offers no protection for the user's location privacy. Therefore, the challenge is in controlling the mobility of the chaffs to \emph{maximally resemble the real service while minimally co-locating with the real service}.
To address this challenge, we study the following closely related questions: (i) How will an eavesdropper track a user in the presence of chaffs? (ii) How should the user control the chaffs to defend against the eavesdropper? (iii) What if the user's defense mechanism is known to the eavesdropper?

\subsection{Related Work}

Most existing work on location privacy refers to protecting the \emph{direct channel}, where the user intentionally releases his location to access LBS \cite{Gruteser03MobiSys}. Most existing solutions, e.g., \cite{Gruteser03MobiSys,Meyerowitz09MobiCom,Chow11GeoInformatica,Gkoulalas-Divanis10SIGKDD}, use location transformations such as spatial/temporal cloaking to satisfy a given anonymity requirement (e.g., $k$-anonymity). 
The basic idea is to let a trusted server ``cloak'' a user by replacing the exact user locations by bounding boxes containing sufficiently many other users. While such a strategy can protect the direct channel where the release of location information is explicit, it does not protect side channels such as the cyber side channel considered here.

Besides the direct channel, \emph{side channels} can also release location information. An important side channel in wireless networks is the transmission activity, which can be monitored by a wireless eavesdropper to track the user. To defend against wireless eavesdropping, mechanisms are proposed to protect senders/receivers using anonymous routing protocols, frequently changing pseudonyms, silent periods, and reduced transmission power \cite{Hu05SIGCOMM,Jiang07MobiSys}. The above mechanisms are \emph{intrusive} in that they modify the user's behavior. In contrast, we study another side channel arising in MECs due to correlated user mobility and service mobility, and propose a \emph{non-intrusive} defense mechanism using chaffs.

The idea of using chaffs to protect user security/privacy has been explored in other contexts. In communication networks, \cite{Jiang01INFOCOM} uses dummy packets as chaffs to hide the true traffic rates, and \cite{He09Allerton} furthers the idea to hide the transmission patterns of multi-hop flows. 
In cloud computing, \cite{Stolfo12SPW} proposes to use decoy data to protect the real data during a data theft attack. Similarly, \cite{Kontaxis13bookchapter} proposes to use decoy applications running on fake inputs to confuse an insider attacker. However, we are the first to study the use of chaff services to protect user location privacy.
Besides the novel application context, our problem also requires new methodology. Specifically, as a real service needs to migrate dynamically to follow a mobile user, its mobility pattern (in addition to its content) can be used to identify the service. To effectively protect the user, the chaff services have to resemble the real service in both content and mobility. \looseness=-1

Another line of related work is service migrations in MECs. Service migrations in MECs are primarily driven by the need to keep a service close to its user, where the decision on whether to migrate a service depends on both the migration cost (if migrating the service to follow the user) and the communication cost (if serving the user from the original location as the user moves away). Modeling the user's mobility as a \emph{Markov chain (MC)}, several solutions based on \emph{Markov Decision Processes (MDPs)} have been proposed to minimize the total cost under 1-D \cite{Ksentini14ICC,Wang14MILCOM} or 2-D mobility models \cite{Wang15Networking,Taleb16TCC}. 
Here we consider the worst case (in terms of location privacy) that the real service \emph{always} follows the user, and focus on protecting the user's location privacy using chaffs. We leave the study of privacy-aware service migration to future work. 

\subsection{Summary of Contributions}

We consider the problem of protecting user location privacy using chaffs. Our contributions are:

1) We model the eavesdropper as a \emph{maximum likelihood (ML)} detector that aims at detecting the user's trajectory based on multiple observed trajectories.

2) We propose a suite of increasingly sophisticated chaff control strategies for the user: (i) an \emph{impersonating (IM)} strategy that mimics the user's mobility, (ii) an ML strategy that maximizes the likelihood of the chaff's trajectory to mislead the detector, (iii) an \emph{optimal offline (OO)} strategy that minimizes the eavesdropper's tracking accuracy based on the user's entire trajectory, and (iv) an \emph{optimal online} strategy that minimizes the expected tracking accuracy based on the user's past trajectory. We show that strategies (i-iii) can be computed in polynomial time. While strategy (iv) is difficult to compute, we propose an alternative \emph{myopic online (MO)} strategy that is easily computable.

3) Our analysis shows that while the eavesdropper's tracking accuracy is always non-zero under the IM or ML strategy, it may decay to zero under the OO or MO strategy, where we characterize the condition and the decay rate. \looseness=-1

4) We further analyze the robustness of the proposed strategies against an eavesdropper aware of the strategy.
We show that while the deterministic strategies (ML, OO, MO) are vulnerable to such an eavesdropper, their robustness can be improved through simple extensions using randomization. \looseness=-1

5) We evaluate the proposed strategies on both synthetic and real mobility traces. Our evaluations show that beside the chaff control strategy, the user's mobility model also has a significant impact on the tracking accuracy. Nevertheless, our strategies (especially OO and MO) can significantly reduce the tracking accuracy even for users with highly predictable mobility, and the same holds for the randomized strategies even if the strategy is known to the eavesdropper.

\vspace{.5em} The rest of the paper is organized as follows. Section~\ref{sec:Problem Formulation} formulates the problem. Section~\ref{sec:Eavesdropper's Strategy} specifies the model for the eavesdropper. Section~\ref{sec:Chaff Control Strategies} presents chaff control strategies for the user, whose effectiveness is analyzed in Section~\ref{sec:Analysis of Chaff Control Strategies}. Section~\ref{sec:Adanved Eavesdropper} analyzes the robustness of the proposed strategies and proposes amendments. Section~\ref{sec:Performance Evaluation} evaluates the performance through simulations. Then Section~\ref{sec:Conclusion} concludes the paper.
\emph{All the proofs are provided in the appendix.}

\section{Problem Formulation}\label{sec:Problem Formulation}

\subsection{Network Model}\label{subsec:Network Model}

Given a network field deployed with multiple MECs, we quantize the space into \emph{cells} such that each cell corresponds to the coverage area of one MEC. 
Let $\Lc$ denote the set of cells, which also specifies the set of possible user locations from the perspective of a cyber eavesdropper; let $L:=|\Lc|$. Suppose that there is a user of interest running a delay-sensitive service (e.g., augmented reality) that must be co-located with the user. We consider delay-sensitive service as it has been identified as one of the most promising applications in future wireless networks \cite{Banerjee13Report}, while establishing the worst case for location privacy. We leave the study of more flexible services to future work.
Note that although our analysis focuses on the single-user scenario, our solution can be independently applied to protect multiple users in a multi-user scenario, where our results provide performance lower bounds as other coexisting users (and their chaffs) offer additional protection.

\subsection{Eavesdropper and Chaffs}\label{subsec:Eavesdropper Model}

We consider a cyber eavesdropper that observes the trajectories of services as they migrate among the MECs. Such an eavesdropper can be a hacker inside the MEC system, or an untrusted MEC provider (a.k.a. \emph{edge operator} \cite{Schuster16}) that operates the MECs. Under the assumption of delay-sensitive services as in Section~\ref{subsec:Network Model}, the eavesdropper can track the user by detecting the trajectory of his service.

To prevent detection, the user generates $N-1$ ($N>1$) additional trajectories using chaff services. Each chaff service is an independent instance of the same service that the user is accessing, thus indistinguishable from the real service in content. 
The chaff services will consume MEC resources, and the cost incurred by these services is the responsibility of the user. In this regard, \emph{the parameter $N$ captures the user's budget for running chaff services.}
With assistance of the network provider, the user can make a chaff service follow an arbitrary trajectory by sending fake service requests and migration requests to the corresponding MECs, which cause the chaff service to be instantiated or migrated. Alternatively, the service provider can send these requests on behalf of the user.
For example, the service provider can offer chaff-based protection of user location privacy as a service option, and if a user wants the protection, he can choose this option and pay an extra cost to the service provider, who will then run chaff services at selected MECs according to a chaff control strategy.
Since for a cyber eavesdropper, tracking a user is equivalent to tracking his service, we simply refer to the user's service as ``the user'' and the chaff services as ``the chaffs''.

\subsection{Mobility Model}\label{subsec:Mobility Model}

Assume that the user follows a discrete-time ergodic \emph{Markovian chain (MC)} as in \cite{Ksentini14ICC,Wang14MILCOM,Wang15Networking}, with transition matrix $P=(P(x_t|x_{t-1}))_{x_t,x_{t-1}\in \Lc}$. Let $\pi := (\pi(x))_{x\in\Lc}$ denote his steady-state distribution. Assume that $\pi(x)>0$ for all $x\in\Lc$. 
Mobility of the chaffs (i.e., migration of chaff services) is controlled by the user and will be studied later.

For each $u=1,\ldots,N$, let $x_{u,t}\in \Lc$ denote the location of the $u$-th service in time slot $t$, and $\xbf_u :=(x_{u,t})_{t=1}^T$ the trajectory over $T$ slots. Here $u=1$ corresponds to the user, $u=2,\ldots,N$ correspond to the chaffs, and $T\geq 1$ represents the duration of the user's service.

\subsection{Location Privacy in the Presence of Chaffs}

Our goal is to understand the efficacy of protecting user location privacy using chaffs. We achieve this by studying two closely-related problems:

(i) From the eavesdropper's perspective: Given $N$ trajectories generated by a user and $N-1$ chaffs, which trajectory belongs to the user?

(ii) From the user's perspective: Given $N-1$ chaffs, what trajectories should the chaffs follow to cause the worst performance for the eavesdropper? 

We measure the eavesdropper's performance by his \emph{tracking accuracy}, defined as the time-average probability of correctly tracking the user, i.e., if the eavesdropper believes that the $u$-th trajectory belongs to the user, then his tracking accuracy equals ${1\over T} \sum_{t=1}^T \Pr\{x_{u,t} = x_{1,t}\}$. Note that this is different from the detection accuracy, as $u=1$ is sufficient but not necessary for $x_{u,t} = x_{1,t}$.

\section{Eavesdropper's Strategy}\label{sec:Eavesdropper's Strategy}

Given multiple trajectories $\xbf_u := (x_{u,t})_{t=1}^T$ ($u=1,\ldots,N$), the eavesdropper wants to determine which trajectory belongs to the user of interest. We consider a sophisticated eavesdropper who knows the user's mobility model, i.e., the transition matrix $P$. For example, the eavesdropper can obtain this information by profiling how typical users move in the network field. At this point, we assume that the eavesdropper does not know the user's chaff control strategy; this assumption will be revised in Section~\ref{sec:Adanved Eavesdropper}.

Intuitively, the eavesdropper should pick the trajectory that best matches the user's mobility model. Mathematically, this is the trajectory that has the \emph{maximum likelihood (ML)} among all the trajectories.
Under the assumption that all the trajectories have equal prior probability of belonging to the user, the ML trajectory has the maximum posterior probability of belonging to the user. 
Under the Makovian user mobility model in Section~\ref{subsec:Mobility Model}, the ML detector is given by ($[N]:=\{1,\ldots,N\}$)
\begin{align}
&\hspace{-.5em} \uML \hspace{-.25em} = \hspace{-.25em} \argmax_{u\in [N]} p(\xbf_u)  \hspace{-.25em} =  \hspace{-.25em} \argmax_{u\in [N]} \pi(x_{u,1}) \hspace{-.25em} \prod_{t=2}^T  \hspace{-.25em} P(x_{u,t}|x_{u,t-1}). \label{eq:ML - single-user}
\end{align}
The optimization in (\ref{eq:ML - single-user}) can be easily solved in $O(NT)$ time.

\emph{Remark:} In a multi-user scenario, the detector (\ref{eq:ML - single-user}) can also be used to detect a particular user of interest among multiple users, assuming that only the mobility model of the user of interest is known.

\section{User's Strategy}\label{sec:Chaff Control Strategies}

The problem faced by the user is that given $N-1$ chaffs, how to control the mobility of the chaffs, i.e., how to generate the trajectories $\xbf_u$ ($u=2,\ldots,N$), to maximally confuse the eavesdropper.
Depending on the precise definition of ``confusion'', we have the following chaff control strategies.

\subsection{Impersonating Strategy}\label{subsec:Impersonating Strategy}

If the eavesdropper's strategy is unknown, a safe choice for the user is to make the chaffs appear similar to himself, a strategy referred to as the \emph{impersonating (IM) strategy}. Under Markovian user mobility, this strategy makes each chaff follow a trajectory generated independently from the same transition matrix $P$ as followed by the user, which naturally mimics the user's mobility. Under this strategy, all the $N$ trajectories are statistically identical, and therefore any detector, including the ML detector (\ref{eq:ML - single-user}), can only make a random guess. \looseness=-1

\emph{Remark:} From the eavesdropper's perspective, this is the same as a multi-user scenario where all the users follow the same mobility model.

\subsection{Maximum Likelihood Strategy}\label{subsec:ML Strategy}

\subsubsection{The Strategy}

If the user knows that the eavesdropper uses the ML detector (\ref{eq:ML - single-user}), then he can design trajectories for the chaffs to intentionally mislead the detector. A chaff's trajectory can mislead the ML detector only if its likelihood (based on the user's mobility model) is no smaller than the likelihood of the user's trajectory. Since the detector is deterministic, it suffices to use a single chaff as at most one chaff (the one with the ML trajectory) will have effect even if multiple chaffs are used. \looseness=-1

This idea inspires a strategy referred to as the \emph{maximum likelihood (ML) strategy}. Letting $\Lc^T$ denote all possible trajectories of length $T$, this strategy controls the chaff to follow a trajectory $\xbf_2$ that achieves the following optimization:
\begin{align}
\xbf_2 = \argmax_{\xbf\in\Lc^T} p(\xbf) = \argmax_{\xbf\in\Lc^T} \pi(x_1)\prod_{t=2}^T P(x_t|x_{t-1}). \label{eq:Unconstrained ML Chaff}
\end{align}

\subsubsection{The Algorithm}

While the space of all possible trajectories ($\Lc^T$) is too large to explore exhaustively, the optimization problem in (\ref{eq:Unconstrained ML Chaff}) has a physical interpretation that allows a more efficient solution. We will show that problem (\ref{eq:Unconstrained ML Chaff}) can be converted to a \emph{shortest-path problem} as follows.

\begin{figure}[tb]
\vspace{-.5em}
\centering
\includegraphics[width=2.5in]{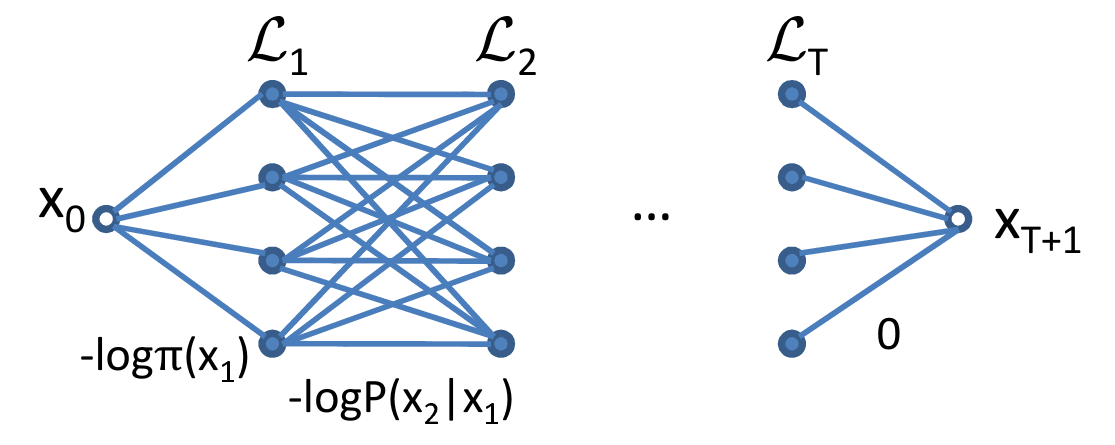}
\vspace{-1em}
\caption{Auxiliary graph for computing the ML trajectory. } \label{fig:unconstrained ML chaff}
\vspace{-.5em}
\end{figure}

The key is to rewrite the optimization (\ref{eq:Unconstrained ML Chaff}) as
\begin{align}\label{eq:Unconstrained ML Chaff equivalent}
\xbf_2 = \argmin_{\xbf\in\Lc^T} -\log{\pi(x_1)} + \sum_{t=2}^T (-\log{P(x_t|x_{t-1})}).
\end{align}
Let $\Lc_t$ ($t=1,\ldots,T$) be a set of vertices representing all possible chaff locations at time $t$ ($|\Lc_t|=|\Lc|$). As illustrated in Fig.~\ref{fig:unconstrained ML chaff}, we construct a graph $\Gc=(V,E)$, with vertices $V=\{x_0\}\cup \{x_{T+1}\}\cup \bigcup_{t=1}^T \Lc_t$ denoting possible chaff locations at different times ($x_0$ and $x_{T+1}$ are virtual locations) and edges $E=(\{x_0\}\times \Lc_1)\cup(\Lc_T\times\{x_{T+1}\})\cup \bigcup_{t=2}^T (\Lc_{t-1}\times\Lc_t)$ denoting possible movements. We assign each edge a cost\footnote{Strictly, each vertex $v\in \Lc_t$ corresponds to a unique cell $f_t(v)\in \Lc$. Edge $(x_0,x)$ for each $x\in\Lc_1$ has cost $-\log\pi(f_1(x))$; edge $(x,x')$ for each $x\in \Lc_{t-1}$ and $x'\in \Lc_t$ has cost $-\log P(f_t(x')|f_{t-1}(x))$ ($t=2,\ldots,T$). }:
\begin{enumerate}
\item edge $(x_0, x)$ for each $x\in\Lc_1$ has cost $-\log\pi(x)$;
\item edge $(x,x')$ for each $x\in \Lc_{t-1}$ and $x'\in \Lc_t$ ($t=2,\ldots,T$) has cost $-\log P(x'|x)$;
\item edge $(x,x_{T+1})$ for each $x\in\Lc_T$ has zero cost.
\end{enumerate}
Each possible trajectory $\xbf = (x_t)_{t=1}^T$ corresponds to a path $(x_0, x_1,\ldots,x_T, x_{T+1})$ from $x_0$ to $x_{T+1}$ in $\mathcal{G}$, and the cost of this path, given by the sum of its edge costs, equals the value of the objective function (\ref{eq:Unconstrained ML Chaff equivalent}) at $\xbf$.
Thus the solution to (\ref{eq:Unconstrained ML Chaff equivalent}) is essentially the path from $x_0$ to $x_{T+1}$ that has the \emph{minimum cost}, which can be computed by Dijkstra's algorithm\footnote{Dijkstra's algorithm works in this case since all the edge costs are non-negative. } at complexity $O(T L^2)$.
Note that this trajectory only depends on the user's mobility model and can thus be computed beforehand.

\emph{Remark:} The ML strategy is clearly optimal against the ML detector (\ref{eq:ML - single-user}) in minimizing the detection accuracy. This is, however, different from minimizing the tracking accuracy, as the chaff's trajectory may coincide with the user's trajectory at times, when the eavesdropper can track the user perfectly.

\subsection{Optimal Offline Strategy}\label{subsec:Optimal Offline Strategy}

\subsubsection{The Strategy}

The ultimate goal of the user is to prevent himself from being tracked by the eavesdropper. To this end, the chaff's trajectory not only needs to mislead the detector, but also needs to be as disjoint as possible from the user's trajectory. For the ML detector (\ref{eq:ML - single-user}), the optimal strategy is to let the chaff follow a trajectory that is as disjoint as possible from the user's trajectory while having a higher likelihood, i.e., the solution $\xbf_2 := (x_{2,t})_{t=1}^T$ to the following optimization\footnote{Here $\mathds{1}_{\{\cdot\}}$ is the indicator function.}
\begin{align}
& \min \sum_{t=1}^T \mathds{1}_{\{x_{2,t}=x_{1,t}\}} \label{eq:min accuracy} \\
\mbox{s.t.} & \, \pi(x_{2,1}) \hspace{-.25em} \prod_{t=2}^T \hspace{-.25em}  P(x_{2,t}|x_{2,t-1})  \hspace{-.25em} >  \hspace{-.25em} \pi(x_{1,1}) \hspace{-.25em} \prod_{t=2}^T \hspace{-.25em}  P(x_{1,t}|x_{1,t-1}), \label{eq:higher likelihood}
\end{align}
where the constraint (\ref{eq:higher likelihood}) guarantees that the ML detector will pick the chaff's trajectory, and the objective (\ref{eq:min accuracy}) minimizes the number of times that the chaff's trajectory coincides with the user's trajectory. Again, a single chaff suffices as the detector is deterministic. We refer to this strategy as the \emph{optimal offline (OO) strategy}, as it is optimal in minimizing the tracking accuracy of an eavesdropper using the ML detector (\ref{eq:ML - single-user}) and it requires knowledge of the entire trajectory (including the future trajectory) of the user.

Note that (\ref{eq:higher likelihood}) will be infeasible if the user's trajectory has the maximum likelihood among all the trajectories. In this case, we change the ``$>$'' in (\ref{eq:higher likelihood}) to ``$=$'' to force the ML detector to make a random guess, but the objective (\ref{eq:min accuracy}) remains valid as we want to minimize the number of times the eavesdropper tracks the user correctly when the detector guesses wrong.

\subsubsection{The Algorithm}\label{subsubsec:Algorithm Design - OO}

While a brute-force solution to (\ref{eq:min accuracy}) is infeasible due to the exponentially large solution space, we can solve it by dynamic programming over the weighted graph introduced in Fig.~\ref{fig:unconstrained ML chaff}. Let $p_{\xbf_1}$ denote the path in this graph corresponding to the user's trajectory, and $K(p_{\xbf_1})$ the length (sum of edge costs) of this path. Then optimizing (\ref{eq:min accuracy}) subject to (\ref{eq:higher likelihood}) is equivalent to finding a path from $x_0$ to $x_{T+1}$ with a length less than $K(p_{\xbf_1})$ (or equal to $K(p_{\xbf_1})$ if $p_{\xbf_1}$ is a shortest path) that is as disjoint as possible from $p_{\xbf_1}$. To this end, we introduce $K_t(x,i)$ to denote the length of the shortest path from $x\in\Lc_t$ to $x_{T+1}$ that intersects (i.e., sharing vertices) with $p_{\xbf_1}$ at most $i$ times ($0\leq i\leq T-t+1$), and $n_t(x,i)$ to denote the next hop neighbor of $x$ on this path.

Initially, $K_T(x,1)\equiv 0$ for all $x\in\Lc_T$,
\begin{align}
K_T(x,0) &= \left\{\begin{array}{ll}
0 &\mbox{if } x\neq x_{1,T},\\
\infty &\mbox{o.w.,}
\end{array} \right.
\end{align}
and $n_T(x,i)\equiv x_{T+1}$ for all $x\in\Lc_T$ and $i\in\{0,\: 1\}$. For $t=T-1,\ldots,1$,
\begin{align}
K_t(x,i)  \hspace{-.25em}  &= \hspace{-.25em} \left\{  \hspace{-.5em} \begin{array}{ll}
 \hspace{-.25em} \min\limits_{x'\in\Lc_{t+1}} \hspace{-.75em}  - \hspace{-.25em} \log P(x'|x) \hspace{-.25em} + \hspace{-.25em} K_{t+1}(x',i) &  \hspace{-2.5em} \mbox{if }x\neq x_{1,t},\\
 \hspace{-.25em} \min\limits_{x'\in\Lc_{t+1}} \hspace{-.75em}  - \hspace{-.25em} \log P(x'|x)  \hspace{-.25em} + \hspace{-.25em}  K_{t+1}(x',i-1) &  \hspace{-.5em} \mbox{o.w.,}
\end{array}
 \right. \nonumber\\
 & \hspace{6em} \forall x\in\Lc_t,\: i\in\{0,\ldots,T-t+1\},
\end{align}
and $n_t(x,i)$ is the value of $x'\in\Lc_{t+1}$ achieving the minimum. By definition, $K_t(x,i)\equiv K_t(x,T-t+1)$ for all $i> T-t+1$, and $K_t(x,i)=\infty$ for $i<0$ (infeasible).
At $t=0$, we have
\begin{align}
K_0(x_0,i)  \hspace{-.25em} &=  \hspace{-.25em} \min_{x\in\Lc_1} \hspace{-.05em}  - \hspace{-.25em} \log \pi(x) \hspace{-.05em}  + \hspace{-.05em}  K_1(x,i),~\forall i \hspace{-.25em} \in \hspace{-.25em} \{0,\ldots,T\},
\end{align}
and $n_0(x_0,i)$ is the $x\in\Lc_1$ achieving the minimum.

Then $i^*$, defined by the smallest value of $i\in \{0,\ldots,T\}$ satisfying $K_0(x_0,i)< K(p_{\xbf_1})$, is the optimal value of (\ref{eq:min accuracy}) (if infeasible, then $i^*$ is the smallest $i$ satisfying $K_0(x_0,i) = K(p_{\xbf_1})$). The optimal chaff's trajectory $\xbf_2$ is given by:\begin{enumerate}
\item $x_{2,1}= n_0(x_0,i^*)$, and $i_1=i^*$;
\item for $t=2,\ldots,T$: $x_{2,t}=n_{t-1}(x_{2,t-1},i_{t-1})$, and $i_t= i_{t-1}$ if $x_{2,t-1}\neq x_{1,t-1}$ or $i_t=i_{t-1}-1$ otherwise.
\end{enumerate}
See Algorithm~\ref{Alg:OO} for the pseudo code of this strategy.
The complexity of this dynamic programming is $O(T^2 L^2)$.

\begin{algorithm}[th!]
\small
\caption{Optimal Offline (OO) Strategy} \label{Alg:OO}
\SetKwInOut{Input}{input}\SetKwInOut{Output}{output}
\Input{Space $\mathcal{L}$, user transition matrix $P$, user steady state distribution $\pi$, time horizon $T$, user trajectory $\mathbf{x}_1$ }
\Output{Chaff trajectory $\mathbf{x}_2=(x_{2,t})_{t=1}^T$}
$K(p_{\mathbf{x}_1}) = -\log{\pi(x_{1,1})}$\;
\ForEach{$t=2,\ldots,T$}
{$K(p_{\mathbf{x}_1}) = K(p_{\mathbf{x}_1}) -\log{P(x_{1,t}|x_{1,t-1})}$}
\ForEach{$x\in\mathcal{L}$}
{$K_T(x,1)=0$\;
\If{$x\neq x_{1,T}$}
{$K_T(x,0)=0$\;}
\Else
{$K_T(x,0)=\infty$\;}
$n_T(x,0)=x_{T+1}$\;
$n_T(x,1)=x_{T+1}$\;
}
\ForEach{$t=T-1,\ldots,1$}
{\ForEach{$x\in \mathcal{L}$}
{
\ForEach{$i=0,\ldots,T-t+1$}
{
\If{$x\neq x_{1,t}$}
{$j=i$\;}
\Else
{$j=i-1$\;}
$K_t(x,i)= \min_{x'\in\mathcal{L}} -\log{P(x'|x)}+K_{t+1}(x',j)$\;
$n_t(x,i) = \argmin_{x'\in\mathcal{L}} -\log{P(x'|x)}+K_{t+1}(x',j)$\;
} 
}
}
\ForEach{$i=0,\ldots,T$}
{$K_0(x_0,i) = \min_{x\in\mathcal{L}} -\log{\pi(x)} + K_1(x,i)$\;
$n_0(x_0,i) = \argmin_{x\in\mathcal{L}} -\log{\pi(x)} + K_1(x,i)$\;
}
\If{$K_0(x_0,T) < K(p_{\mathbf{x}_1})$}
{\ForEach{$i^*=0,\ldots,T$}
{\If{$K_0(x_0,i^*) < K(p_{\mathbf{x}_1})$}
{break\;}
}
}
\Else
{\ForEach{$i^*=0,\ldots,T$}
{\If{$K_0(x_0,i^*) = K(p_{\mathbf{x}_1})$}
{break\;}
}
}
$x_{2,1}=n_0(x_0,i^*)$\; $i_1=i^*$\;
\ForEach{$t=2,\ldots,T$}
{$x_{2,t}=n_{t-1}(x_{2,t-1},i_{t-1})$\;
\If{$x_{2,t-1}\neq x_{1,t-1}$}
{$i_t=i_{t-1}$}
\Else
{$i_t = i_{t-1}-1$}
}
\vspace{-.05em}
\normalsize
\end{algorithm}

\subsection{Optimal Online Strategy}\label{subsec:Online Strategy}

\subsubsection{The Strategy}

In cases where the user's future trajectory cannot be exactly predicted beforehand, the offline strategy is not applicable. For such cases, we consider the online counterpart of the optimization (\ref{eq:min accuracy}), which only requires knowledge of the user's past trajectory and the transition probabilities of the user mobility model (defined in Section~\ref{subsec:Mobility Model}). As shown below, this problem can be cast as a finite-horizon \emph{Markov Decision Process (MDP)}, which is characterized by a $5$-tuple $(\mathcal{S},\: \mathcal{A},\: \mathcal{T},\: C,\: T)$, defined as:

$\bullet$ The state space $\mathcal{S}=\mathbb{R}\times \Lc^2$ is the space of the triple $(\gamma_t,x_{1,t},x_{2,t})$, where $\gamma_t:= \log p(\xbf_1^t)-\log p(\xbf_2^t)$ is the difference between the log-likelihoods of user's/chaff's trajectories ($\xbf_i^t:= (x_{i,1},\ldots,x_{i,t})$), $x_{1,t}$ is the user location, and $x_{2,t}$ is the chaff location, all at time $t$;

$\bullet$ The action space $\mathcal{A}=\Lc$ is the set of possible locations that the chaff can move to at any given time;

$\bullet$ The state transition $\mathcal{T}$ includes three transitions (logically) occurring as: (i) $x_{1,t-1}$ transits to a random $x_{1,t}$ with probability $P(x_{1,t}|x_{1,t-1})$; (ii) $x_{2,t-1}$ transits to a (random or deterministic) $x_{2,t}$ according to a control policy $\psi$; (iii) $\gamma_{t-1}$ transits to $\gamma_t = \gamma_{t-1} + \log P(x_{1,t}|x_{1,t-1}) - \log P(x_{2,t}|x_{2,t-1})$;

$\bullet$ The cost function $C(\gamma_t,x_{1,t},x_{2,t})= \mathds{1}_{\{x_{2,t}=x_{1,t}\}} + \mathds{1}_{\{x_{2,t}\neq x_{1,t}\}}(\mathds{1}_{\{\gamma_t>0\}}+{1\over 2}\mathds{1}_{\{\gamma_t=0\}})$;

$\bullet$ The horizon $T$ is the time duration of the user's trajectory.

Since the cost function $C(\gamma_t,x_{1,t},x_{2,t})$ represents the per-slot tracking accuracy of the ML detector (\ref{eq:ML - single-user}), the control policy $\psi^{\mbox{\tiny OPT}}$ that minimizes the total cost over horizon $T$ is the optimal online chaff control strategy. 

Solving this MDP optimally, however, faces both the usual challenge of dimensionality and an unusual challenge that one component of the state ($\gamma_t$) has a continuous space. Instead, we consider a commonly used heuristic, the myopic policy, which only minimizes the immediate cost:
\begin{align}\label{eq: psiMY}
\psiMY \hspace{-.15em} ( \hspace{-.15em} \gamma_{t-1},x_{1,t-1},x_{2,t-1},x_{1,t} \hspace{-.15em} ) \hspace{-.25em} &:= \hspace{-.25em}  \argmin_{x_{2,t}\in\Lc}  \hspace{-.1em} C(\gamma_t,x_{1,t},x_{2,t}),
\end{align}
where $\gamma_t$ is completely determined by $\gamma_{t-1}$, $x_{1,t-1}$, $x_{2,t-1}$, $x_{1,t}$, and $x_{2,t}$.
However, any efficient MDP solver (e.g., rollout algorithm) is applicable here, and we leave comparison between different solvers to future work.

\subsubsection{The Algorithm}\label{subsubsec:MO}

Based on (\ref{eq: psiMY}), we develop a control strategy called the \emph{myopic online (MO) strategy}. This strategy combines the maximization of the cumulative likelihood and the minimization of the per-slot tracking accuracy: for each $t=1,\ldots,T$, given the user's location $x_{1,t}$,
\begin{enumerate}
\item if the ML location for the chaff $x_{2,t}^{(1)}=\argmax_{x\in\Lc} P(x|x_{2,t-1})$ (or $\argmax_{x\in\Lc} \pi(x)$ if $t=1$) does not coincide with the user's location $x_{1,t}$, then move the chaff to $x_{2,t}^{(1)}$;
\item otherwise, if the second ML location for the chaff $x_{2,t}^{(2)}=\argmax_{x\in\Lc\setminus\{x_{1,t}\}} P(x|x_{2,t-1})$ (or $\argmax_{x\in \Lc\setminus\{x_{1,1}\}} \pi(x)$ if $t=1$) is good enough (i.e., giving an overall likelihood no smaller than that of the user), then move the chaff to $x_{2,t}^{(2)}$;
\item otherwise, move the chaff to $x_{2,t}^{(1)}$.
\end{enumerate}
Note that in the third case, the user will be tracked correctly at $t$ no matter where the chaff moves, and hence we move the chaff to the ML location to maximize the chance of evading tracking in future slots.
See Algorithm~\ref{Alg:Myopic} for the pseudo code of this strategy, where lines~\ref{MO: case 1, t=1} and \ref{MO: case 1, t>1} handle case (1), lines~\ref{MO: case 2, t=1} and \ref{MO: case 2, t>1} handle case (2), and lines~\ref{MO: case 3, t=1} and \ref{MO: case 3, t>1} handle case (3).

\begin{algorithm}[th!]
\small
\caption{Myopic Online (MO) Strategy} \label{Alg:Myopic}
\SetKwInOut{Input}{input}\SetKwInOut{Output}{output}
\Input{Space $\Lc$, user transition matrix $P$, user steady state distribution $\pi$, time horizon $T$ }
\Output{Locations of chaff in slots $1,\ldots,T$}
observe initial user location $x_{1,1}$\; \label{Myopic:observe initial}
compute $x_{2,1}^{(1)}=\argmax_{x\in\Lc} \pi(x)$\; \label{Myopic:compute ML}
\If{$x_{2,1}^{(1)}\neq x_{1,1}$}
{$x_{2,1}\leftarrow x_{2,1}^{(1)}$\; \label{MO: case 1, t=1}}
\Else
{compute $x_{2,1}^{(2)}=\argmax_{x\in \Lc\setminus\{x_{1,1}\}} \pi(x)$\;
\If{$\pi(x_{2,1}^{(2)})\geq \pi(x_{1,1})$}
{$x_{2,1}\leftarrow x_{2,1}^{(2)}$\; \label{MO: case 2, t=1}}
\Else
{$x_{2,1}\leftarrow x_{2,1}^{(1)}$\; \label{MO: case 3, t=1}}
}\label{Myopic:go to ML}
$\gamma_1\leftarrow \log \pi(x_{1,1}) - \log \pi(x_{2,1})$\; \label{Myopic:update gamma_1}
\ForEach{$t=2,\ldots,T$}
{
observe new user location $x_{1,t}$\;\label{Myopic:observe t}
compute $x_{2,t}^{(1)}=\argmax_{x\in\Lc} P(x|x_{2,t-1})$\;
\If{$x_{2,t}^{(1)}\neq x_{1,t}$}
{$x_{2,t}\leftarrow x_{2,t}^{(1)}$\; \label{MO: case 1, t>1}}
\Else
{
compute $x_{2,t}^{(2)}=\argmax_{x\in\Lc\setminus\{x_{1,t}\}} P(x|x_{2,t-1})$\;
\If{$\gamma_{t-1} + \log P(x_{1,t}|x_{1,t-1}) - \log P(x_{2,t}^{(2)}|x_{2,t-1})\leq 0$}
{$x_{2,t}\leftarrow x_{2,t}^{(2)}$\; \label{MO: case 2, t>1}}
\Else
{$x_{2,t}\leftarrow x_{2,t}^{(1)}$\; \label{MO: case 3, t>1}}
}
$\gamma_t \leftarrow \gamma_{t-1}+\log P(x_{1,t}|x_{1,t-1}) - \log P(x_{2,t}|x_{2,t-1})$\; \label{Myopic:update gamma_t}
}
\vspace{-.05em}
\normalsize
\end{algorithm}

\section{Performance Analysis}\label{sec:Analysis of Chaff Control Strategies}

We now analyze the performance of the proposed strategies in Section~\ref{sec:Chaff Control Strategies} in terms of the tracking accuracy of the eavesdropper in Section~\ref{sec:Eavesdropper's Strategy}. We denote the time-average tracking accuracy under each strategy by $\PIM$, $\PML$, $\POO$, and $\PMO$.

\subsection{Tracking Accuracy under IM}\label{subsec:Accuracy - Impersonating Strategy}

Under the IM strategy, the eavesdropper randomly guesses a trajectory for the user. He correctly tracks the user at time $t$ if and only if (i) he guesses the trajectory right, which occurs with probability $1/N$, or (ii) he guesses the trajectory wrong but the guessed trajectory coincides with the user's trajectory at time $t$. Thus, the overall tracking accuracy equals
\begin{align}
\PIM &= {1\over N} + {N-1\over N}\cdot {1\over T}\sum_{t=1}^T \Pr\{x'_t = x_t\}, \label{eq:PUI - 1}
\end{align}
where $\xbf'=(x'_t)_{t=1}^T$ and $\xbf=(x_t)_{t=1}^T$ are two independent instances of the same MC that describes the user's mobility. Given the steady-state distribution $\pi$ of this MC, it is easy to see that $\Pr\{x'_t = x_t\} = \sum_{x\in\Lc}\pi^2(x)$. Therefore,
\begin{align}
\PIM &= \Big(\sum_{x\in\Lc}\pi^2(x) \Big) + {1\over N}\Big(1-\sum_{x\in\Lc}\pi^2(x) \Big). \label{eq:PUI}
\end{align}

\emph{Remark:} (i) As the number of chaffs increases, the tracking accuracy under the IM strategy converges monotonically at rate $O(1/N)$. (ii) The limit $\lim_{N\to\infty}\PIM = \sum_{x\in\Lc}\pi^2(x)\geq 1/L$, where the lower bound is achieved if and only if $\pi$ is a uniform distribution. Thus, under the IM strategy, the tracking accuracy is bounded away from zero even with infinite chaffs. 

\subsection{Tracking Accuracy under ML}\label{subsec:Accuracy - ML Strategy}

Under the ML strategy, the chaff's trajectory $\xbf_2$ is deterministic and is guaranteed to be selected by the ML detector\footnote{We ignore ties as they occur with an exponentially decaying probability (except for i.i.d. uniform mobility). }. The tracking accuracy is therefore determined by the fraction of time that the user's trajectory coincides with $\xbf_2$, i.e.,
\begin{align}
\PML &= {1\over T}\sum_{t=1}^T \Pr\{x_{1,t} = x_{2,t}\} = {1\over T}\sum_{t=1}^T \pi(x_{2,t}), \label{eq:PUM}
\end{align}
where $\xbf_2$ is the solution to (\ref{eq:Unconstrained ML Chaff}). 

Compared with the value of $\lim_{N\to\infty}\PIM$, $\PML$ can be either smaller or larger. However, we have a fixed comparison when the following lemma applies. 

\begin{lemma}\label{lem:PUI vs PUM}
For any distribution $(\pi(x))_{x\in\Lc}$, $\sum_{x\in\Lc} \pi^2(x) \leq \max_{x\in\Lc}\pi(x)$, and ``$=$'' holds if and only if $(\pi(x))_{x\in\Lc}$ is a uniform distribution.
\end{lemma}

\emph{Remark:} Therefore, if the ML chaff always stays in the same cell (the cell with the maximum steady-state probability), then it is better to use a sufficiently large number of IM chaffs.

\subsection{Tracking Accuracy under OO}\label{subsec:Accuracy - OO Strategy}

Under the OO strategy, the chaff's trajectory is designed to yield the minimum tracking accuracy. Therefore, its tracking accuracy is upper-bounded by the tracking accuracy under any suboptimal strategy.

\subsubsection{Auxiliary Strategy}

To bound the tracking accuracy under the OO strategy, we introduce a suboptimal strategy whose tracking accuracy can be analyzed in closed form. This strategy, referred to as the \emph{constrained maximum likelihood (CML) strategy}, greedily maximizes the likelihood of the chaff's trajectory under the constraint that the chaff cannot co-locate with the user. That is, given the user's trajectory $\xbf_1$, the chaff's trajectory $\xbf_2$ is computed by \begin{enumerate}
\item at $t=1$, $x_{2,1} = \argmax_{x\in \Lc\setminus \{x_{1,1}\}} \pi(x)$;
\item at $t>1$, $x_{2,t} = \argmax_{x\in\Lc \setminus \{x_{1,t}\}} P(x|x_{2,t-1})$.
\end{enumerate}
Note that CML is actually an online strategy as it never requires the future trajectory of the user.

\subsubsection{Analysis of Auxiliary Strategy}\label{subsubsec:Analysis of CML}

Under the CML strategy, the chaff's trajectory is always disjoint from the user's trajectory, and thus the eavesdropper correctly tracks the user if and only if the ML detector is correct, which occurs only if the user's trajectory has a likelihood no smaller than that of the chaff's trajectory. That is, the tracking accuracy under the CML strategy satisfies
\begin{align}
\PCML \leq \Pr\{p(\xbf_1)\geq p(\xbf_2)\},
\end{align}
where $\xbf_2$ is generated according to the CML strategy.

As $p(\xbf) = \pi(x_1)\prod_{t=2}^T P(x_t|x_{t-1})$, we can define
\begin{align}
& c_1(x_{1,1}, x_{2,1}) := \log{\pi(x_{1,1})} \hspace{-.05em}-\hspace{-.05em} \log{\pi(x_{2,1})}, \label{eq:c_1}\\
&\hspace{-1em} c_t(x_{1,t}, x_{2,t}, x_{1,t-1}, x_{2,t-1}) := \log{P(x_{1,t}|x_{1,t-1})} \nonumber\\
&\hspace{8em} - \log{P(x_{2,t}|x_{2,t-1})}, ~~~ t>1, \label{eq:c_t}
\end{align}
and convert $\Pr\{p(\xbf_1)\geq p(\xbf_2)\}$ to
\begin{align}
&\hspace{-.75em} \Pr\{c_1(x_{1,1}, x_{2,1}) \hspace{-.25em} + \hspace{-.25em} \sum_{t=2}^T c_t(x_{1,t}, x_{2,t}, x_{1,t-1}, x_{2,t-1}) \geq 0\}. \label{eq:P(sum of c_t >= 0)}
\end{align}
The tracking accuracy under the CML strategy is then upper-bounded by (\ref{eq:P(sum of c_t >= 0)}).

To bound (\ref{eq:P(sum of c_t >= 0)}), we consider a new MC formed by $y_t:= (x_{1,t},\: x_{2,t})$. This MC is induced by the original MC describing the user's mobility, with a transition probability
\begin{align}
&\hspace{-1em} P(y_t|y_{t-1}) \hspace{-.25em} = \hspace{-.25em} \left\{\hspace{-.5em}\begin{array}{ll}
P(x_{1,t}|x_{1,t-1}) & \hspace{-.5em}\mbox{if } x_{2,t} \hspace{-.25em} = \hspace{-.25em} f(x_{1,t},x_{2,t-1}),\\
0 & \mbox{o.w.,}
\end{array} \right. \label{eq:P(y|y') - constrained ML}
\end{align}
where $f(x_{1,t},x_{2,t-1}) := \argmax_{x\in\Lc \setminus \{x_{1,t}\}} P(x|x_{2,t-1})$ is the chaff location at time $t$ under the CML strategy.
With a little abuse of notation, let $\pi(y_t)$ denote the steady-state distribution of the MC $\{y_t\}_{t=1}^\infty$. For any $\epsilon>0$, let $\tmix(\epsilon)$ denote the \emph{$\epsilon$-mixing time} of $\{y_t\}_{t=1}^\infty$ \cite{Levin09book}. For the ease of notation, we shorten $c_t(x_{1,t}, x_{2,t}, x_{1,t-1}, x_{2,t-1})$ to $c_t$.

\begin{figure}[tb]
\vspace{-.5em}
\centering
\includegraphics[width=3in]{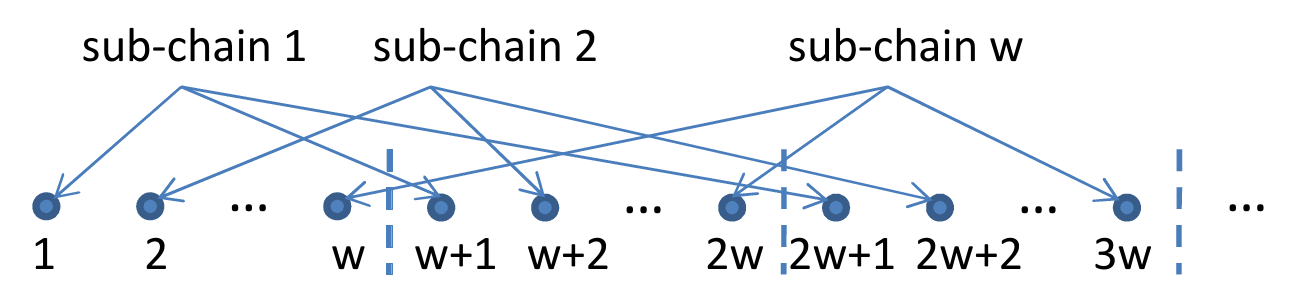}
\vspace{-1em}
\caption{Decomposing a chain into $w$ sub-chains. } \label{fig:MC_decomposition}
\vspace{-.5em}
\end{figure}

Our idea is to apply concentration bounds to show that (\ref{eq:P(sum of c_t >= 0)}) diminishes with $T$ if $\mbbE[c_t]<0$. 
However, since $c_t$ is a function of $y_t$ and $y_{t-1}$, $c_t$'s are correlated, which makes existing concentration bounds inapplicable. To address this challenge, we decompose $\sum_t c_t$ into $w$ summations $\sum_k c_{kw+i}$ ($i=1,\ldots,w$) such that each summation is over a sub-chain consisting of elements that are $w$ steps apart, as illustrated in Fig.~\ref{fig:MC_decomposition}. Intuitively, if $w$ is sufficiently large, correlation within a sub-chain will be sufficiently weak such that the usual concentration bounds hold for summation over the sub-chain.

To formalize this intuition, we prove the following lemma. Define (noting that $c_t$ is determined by $y_t$ and $y_{t-1}$)
\begin{align}\label{eq:g(y)}
g(y_{t-1}) &:= \mbbE[c_t | y_{t-1}] = \hspace{-.75em} \sum_{y_{t}\in \Lc^2}  \hspace{-.5em} P(y_{t}|y_{t-1}) c_t.
\end{align}

\begin{lemma}\label{lem:near-constant mean}
For any $\epsilon>0$, if $w = \tmix(\epsilon)+1$, then
\begin{align}
\Big|\mbbE[c_{kw+i}|c_{k'w+i}, \forall 0\leq k'<k] - \mbbE[c_{kw+i}]\Big| \leq \epsilon \delta
\end{align}
for all $k\geq 0$ and $i\in \{1,\ldots,w\}$, where $\delta := \min(\sum_{y\in \Lc^2}|g(y)|,\: 2 \max_{y\in \Lc^2}|g(y)|)$.
\end{lemma}

To use the result in Lemma~\ref{lem:near-constant mean}, we need a concentration bound for possibly correlated random variables with bounded conditional expectations. We prove the following bound.

\begin{lemma}\label{lem:generalized Chernoff-Hoeffding}
Let $X_1,\ldots,X_n$ be random variables with range $[a,\: b]$, and $\mbbE[X_t|X_1,\ldots,X_{t-1}]\in [\mu-\epsilon,\: \mu]$ for all $1\leq t\leq n$ ($\epsilon>0$). Let $S_n := \sum_{t=1}^n X_t$. For all $\Delta\geq 0$,
\begin{align}
\Pr\{S_n \geq n(\mu+\Delta)\} &\leq e^{-2n\Delta^2/(b-a+\epsilon)^2}.
\end{align}
\end{lemma}

\emph{Remark:} This bound generalizes the \emph{Chernoff-Hoeffding bound} \cite{Pollard84book}, as the original bound requires $X_t\in [0,\: 1]$ and $\mbbE[X_t|X_1,\ldots,X_{t-1}] = \mu$ for all $t$.

We are now ready to bound (\ref{eq:P(sum of c_t >= 0)}). We will need the following constants. Let $c_0$ be the maximum value of $c_1$, and $\cmin$ ($\cmax$) be the minimum (maximum) value of $c_t$ for $t>1$. Specifically, given the user's transition matrix $P$ and steady-state distribution $\pi$, let $\pimax$ and $\pi_2$ denote the largest and the second largest steady-state probabilities, $\pmax$ and $\pmin$ denote the maximum/minimum (positive) transition probabilities, $p_2(x')$ denote the second largest transition probabilities among $\{P(x|x'):\: x\in\Lc\}$, and $p_2 := \min_{x'\in\Lc} p_2(x')$. Then $c_0 = \log(\pimax/\pi_2)$, $\cmin = \log(\pmin/ \pmax)$, and $\cmax = \log(\pmax / p_2)$.

\begin{theorem}\label{thm:accuracy under CML}
Let $\mbbE[c_t] := -\mu$ ($t>1$). If $\exists \epsilon>0$ such that $\mu-\epsilon \delta - c_0/(T-w) \geq 0$ for $w$ and $\delta$ defined in Lemma~\ref{lem:near-constant mean}, then the tracking accuracy under the CML strategy (and thus the OO strategy) satisfies
\begin{align}
\POO \leq \PCML \leq w \cdot \exp\left(-2\Big({T\over w}-1\Big) {(\mu-\epsilon \delta-{c_0\over T-w})^2 \over (\cmax-\cmin +2\epsilon \delta)^2} \right).
\end{align}
\end{theorem}

\emph{Remark:} A few remarks are in order:

i) In contrast to the previous strategies (IM, ML) where the tracking accuracy is always non-zero, we see that when the condition in Theorem~\ref{thm:accuracy under CML} holds, the CML strategy (and hence the OO strategy) can reduce the tracking accuracy to zero exponentially fast in time.

ii) For a sufficiently large $T$, the condition in Theorem~\ref{thm:accuracy under CML} holds if and only if $\mbbE[c_t]<0$, which has an information-theoretic interpretation: By definition, $\mbbE[c_t] = H(X_{2,t}|X_{2,t-1}) - H(X_{1,t}|X_{1,t-1})$, where $H(X_{1,t}|X_{1,t-1})$ ($H(X_{2,t}|X_{2,t-1})$) is the \emph{conditional entropy} of the user's (chaff's) movement. Thus, the tracking accuracy decays to zero if $H(X_{1,t}|X_{1,t-1}) > H(X_{2,t}|X_{2,t-1})$, i.e., the user has a higher entropy than the chaff (under the CML strategy). 

\subsection{Tracking Accuracy under MO}\label{subsec:Accuracy - MO Strategy}

We first analyze the per-slot tracking accuracy under the MO strategy at time $T$, denoted by $\PMO(T)$. Recall that $c_t$ ($t\geq 1$) is the per-slot difference in log-likelihoods defined in (\ref{eq:c_1}, \ref{eq:c_t}), $c_0$ is the maximum value of $c_1$, and $\cmin$ ($\cmax$) is the minimum (maximum) value of $c_t$ for\footnote{As in CML, the MO strategy only moves the chaff to the ML or the second ML location, and thus the values of $c_0$, $\cmin$, and $\cmax$ remain the same as in Theorem~\ref{thm:accuracy under CML}. } $t>1$.

Under the MO strategy, the user is tracked correctly at time $T$ only if $x_{2,T}=x_{1,T}$ or $\gamma_T\geq 0$, where $\gamma_t$ ($t=1,\ldots,T$) is defined in Section~\ref{subsec:Online Strategy}. According to MO, $x_{2,T}=x_{1,T}$ only if $x_{1,T}=x_{2,T}^{(1)}$ and $\gamma_{T-1}+\log P(x_{1,T}|x_{1,T-1}) - \log P(x_{2,T}^{(2)}|x_{2,T-1}) >0$ ($x_{2,T}^{(1)}$ and $x_{2,T}^{(2)}$ are computed as in Section~\ref{subsubsec:MO}), which holds only if $\gamma_{T-1} > - \cmax$. Meanwhile, $\gamma_T\geq 0$ holds only if $\gamma_{T-1}\geq -\cmax$. Therefore,
\begin{align}\label{eq:PMO(T) upper bound, initial}
&\hspace{-1em}\PMO(T) \leq \Pr\{\gamma_{T-1}\geq -\cmax\} = \Pr\{\sum_{t=1}^{T-1}c_t \geq -\cmax\}.
\end{align}

We follow steps similar to Section~\ref{subsubsec:Analysis of CML} to bound (\ref{eq:PMO(T) upper bound, initial}). First, we define a new MC with state $z_t := (\gamma_t, x_{1,t}, x_{2,t})$ and transition probability
\begin{align}\label{eq:P(z|z') - MO}
&\hspace{-.5em}P(z_t|z_{t-1}) \hspace{-.25em}=\hspace{-.25em}\left\{\hspace{-.5em}\begin{array}{ll}
P(x_{1,t}|x_{1,t-1}) &\hspace{-.5em}\mbox{if } x_{2,t}=f_1(z_{t-1},x_{1,t}),\\
 & \hspace{-1em} \gamma_t = f_2(z_{t-1},x_{1,t},x_{2,t}), \\
0 &\mbox{o.w.},
\end{array}
 \right.
\end{align}
where $f_1(z_{t-1},x_{1,t})$ is the chaff's location at time $t$ given by MO for state $(z_{t-1}, x_{1,t})$, and $f_2(z_{t-1},x_{1,t},x_{2,t}) := \gamma_{t-1}+\log P(x_{1,t}|x_{1,t-1})-\log P(x_{2,t}|x_{2,t-1})$. The MC $\{z_t\}_{t=1}^\infty$ captures the system evolution under the MO strategy.
Let $\pi'(z_t)$ denote the steady-state distribution and $\tmix'(\epsilon)$ ($\forall \epsilon>0$) the $\epsilon$-mixing time of $\{z_t\}_{t=1}^\infty$.

Next, define $g'(z)$ as in (\ref{eq:g(y)}) by replacing $P(y_t|y_{t-1})$ by the new transition probability (\ref{eq:P(z|z') - MO}) (note that $c_t$ is determined by $z_t$ and $z_{t-1}$). Lemma~\ref{lem:near-constant mean} still holds, with $w$ replaced by $w' := \tmix'(\epsilon)+1$ and $\delta$ replaced by $\delta' := 2\max_{z\in \mathbb{R}\times\Lc^2}|g'(z)|$.
Arguments similar to Theorem~\ref{thm:accuracy under CML} lead to the following result.

\begin{theorem}\label{thm:accuracy under MO}
Let $\mbbE[c_t] := -\mu'$ ($t>1$) under the MO strategy. If $\exists \epsilon>0$ such that $\mu'-\epsilon\delta'-{c_0+\cmax \over T-w'-1} \geq 0$ for $w'$ and $\delta'$ defined above, then the tracking accuracy at time $T$ under the MO strategy satisfies
\begin{align}
\PMO(T) \leq w'\hspace{-.25em}\cdot\hspace{-.05em} \exp\left(\hspace{-.25em}-2\Big({T-w'-1\over w'}\Big){(\mu'-\epsilon\delta'-{c_0+\cmax\over T-w'-1})^2 \over (\cmax-\cmin+2\epsilon\delta')^2} \hspace{-.25em}\right).
\end{align}
\end{theorem}

\emph{Remark:} As in Theorem~\ref{thm:accuracy under CML}, Theorem~\ref{thm:accuracy under MO} implies that the MO strategy can drive the per-slot tracking accuracy to zero if $\mbbE[c_t]<0$, i.e., the chaff's movement has a lower entropy (or a larger average log-likelihood) than the user's movement, except that now the chaff follows the MO strategy.

Finally, using Theorem~\ref{thm:accuracy under MO}, we can bound the time-average tracking accuracy as follows.

\begin{corollary}\label{coro:time-average accuracy under MO}
Suppose that the condition in Theorem~\ref{thm:accuracy under MO} holds for $T$. Let $T_0\leq T$ be the smallest value for which the condition holds, and
\begin{align}
\alpha &:= {2(\mu'-\epsilon\delta'-{c_0+\cmax\over T_0-w'-1})^2 \over w'(\cmax-\cmin+2\epsilon\delta')^2}.
\end{align}
The overall tracking accuracy under the MO strategy satisfies\looseness=-1
\begin{align}
\PMO &\leq {1\over T}\left(T_0-1+{w'e^{\alpha(w'+1-T_0)}\over 1-e^{-\alpha}} \right).
\end{align}
\end{corollary}

\emph{Remark:} Compared to the exponential decay under the CML (or OO) strategy in Theorem~\ref{thm:accuracy under CML}, we see that the MO strategy yields a slower decay of $O(1/T)$ according to the bound. However, its actual performance is not necessarily worse, as verified through simulations (see Section~\ref{sec:Performance Evaluation}).

\section{Robustness to Advanced Eavesdropper}\label{sec:Adanved Eavesdropper}

We have assumed that the eavesdropper always applies the ML detector (\ref{eq:ML - single-user}) regardless of the chaff control strategy of the user. If he is aware of the user's strategy, however, he may use a different detector. The question is: how robust is a chaff control strategy to an eavesdropper aware of the strategy?

\subsection{Robustness Analysis}

\subsubsection{Robustness of IM}\label{subsubsec:Robustness of Impersonating Chaff}

Under this strategy, each chaff follows a trajectory that is statistically identical to the user's trajectory, and thus the eavesdropper has to randomly guess a trajectory even if the strategy is known to him. Therefore, the IM strategy is fully robust, i.e., an advanced eavesdropper knowing the strategy has the same accuracy as a basic eavesdropper without such knowledge (as in Section~\ref{subsec:Accuracy - Impersonating Strategy}).

\subsubsection{Robustness of ML}\label{subsubsec:Robustness of ML Chaff}

Intuitively, any deterministic strategy will perform poorly if the eavesdropper knows the strategy. Specifically, knowing that the user uses the ML strategy, the eavesdropper can compute the chaff's trajectory according to this strategy (if any tie, suppose the tie breaker is also known) and ignore any observed trajectory that matches the chaff's trajectory. This eavesdropper can always track the user correctly, as the user's trajectory will either coincide with the chaff's trajectory (and be trivially tracked), or deviate from the chaff's trajectory at some point and be detected.

\subsubsection{Robustness of OO and MO}\label{subsubsec:Robustness of OO and MO}

Under the OO/MO strategy, the chaff's trajectory is a deterministic function of the user's trajectory; denote this function by $\Gamma_i(\xbf_1)$, where $i=$ OO for the OO strategy and $i=$ MO for the MO strategy.
Knowing the strategy (and hence $\Gamma_i(\cdot)$), the eavesdropper can compute $\Gamma_i(\xbf)$ for each observed trajectory $\xbf$ and ignore a trajectory $\xbf'\neq \xbf$ if $\xbf' = \Gamma_i(\xbf)$; if both trajectories are ignored, a random guess is made.
This eavesdropper makes a mistake only if $\xbf_1 = \Gamma_i(\xbf_2)$ (the user appears as a ``chaff'' of the chaff), which occurs with an exponentially decaying probability $\pimax \pmax^{T-1}$.

\subsection{Defense against Advanced Eavesdropper}\label{subsec:Defense against Adanced Eavesdropper}

We see that all the strategies, except for IM, are vulnerable if the eavesdropper learns the employed strategy. We can, however, improve their robustness through simple extensions. One idea is to generate multiple trajectories for chaffs by introducing random perturbations to the original strategy.

\subsubsection{Robust ML (RML) Strategy}

We perturb the ML strategy by introducing a set of cell-slot pairs $\Xc_u=\{(l,\: t):\: l\in\Lc,\: t\in [T]\}$ ($[T]:=\{1,\ldots,T\}$) for each $u=2,\ldots,N$ such that trajectory $\xbf_u$ must avoid cell $l$ at time slot $t$ for each $(l,\: t) \in\Xc_u$.
The perturbed strategy, referred to as the \emph{robust ML (RML) strategy}, generates chaffs' trajectories iteratively: for each $u=2,\ldots,N$,
\begin{enumerate}
\item form $\Xc_u$ by randomly select a pair from $\{(x_{u',t},\: t): t\in [T]\}$ for each $u'=1,\ldots,u-1$;
\item compute $\xbf_u$ by solving for the shortest path from $x_0$ to $x_{T+1}$ in $\mathcal{G}-\Xc_u$, which denotes a subgraph of $\mathcal{G}$ defined in Fig.~\ref{fig:unconstrained ML chaff} generated by removing the vertex representing $x$ from $\Lc_t$ for each $(x,\: t)\in\Xc_u$.
\end{enumerate}
The edge costs in $\mathcal{G}$ (see Section~\ref{subsec:ML Strategy}) imply that each $\xbf_u$ constructed as above is an ML trajectory that avoids $\Xc_u$.

\subsubsection{Robust OO (ROO) Strategy}

Following a similar idea, we modify the OO strategy to the \emph{robust OO (ROO) strategy} by following the same iterative process as in RML, except that Step~2) is replaced by a variation of the dynamic programming in Section~\ref{subsubsec:Algorithm Design - OO}, with $\Lc_t$ replaced by $\Lc'_t := \Lc_t\setminus \{x:\: (x,t)\in \Xc_u\}$ ($t\in [T]$).

\subsubsection{Robust MO (RMO) Strategy}

To maintain the online property of the MO strategy, we replace $\Xc_u$ ($u=2,\ldots,N$) by a set of index-slot pairs $\Xc'_u=\{(u',t):\: u'\in [N],\: t\in [T]\}$, where each $(u',t)\in \Xc'_u$ denotes that we want trajectory $\xbf_u$ to avoid trajectory $\xbf_{u'}$ at time $t$. The \emph{robust MO (RMO) strategy} generates $\Xc'_u$ beforehand by $\Xc'_u = \{(u',t_{u'})\}_{u'=1}^{u-1}$, where each $t_{u'}$ is randomly selected from $[T]$.
Then for each $t=1,\ldots,T$, it determines the chaffs' locations sequentially: for each $u=2,\ldots,N$, $x_{u,t}$ is determined as in Section~\ref{subsubsec:MO}, except that $\Lc$ is replaced by $\Lc-\{x_{u',t}:\: (u',t)\in \Xc'_u\}$.

\vspace{.5em}
\emph{Discussion:} The above robust strategies fully utilize all the $N-1$ chaffs and randomize their trajectories to prevent them from being recognized by the eavesdropper. Meanwhile, these strategies also approximate their original versions in terms of the performance under the ML detector (see Section~\ref{sec:Performance Evaluation}).

\section{Performance Evaluation}\label{sec:Performance Evaluation}

We use both synthetic and trace-driven simulations to evaluate the effectiveness of the proposed chaff control strategies. We measure the effectiveness of a chaff control strategy by the eavesdropper's tracking accuracy; the lower the accuracy, the more effective the strategy.

\subsection{Synthetic Simulations}\label{subsec:Synthetic Simulations}

\subsubsection{Simulation Setting}\label{subsec:Simulation Setting}

We generate synthetic mobility traces, where the user follows a MC of $L$ states with transition probabilities specified below, and the chaffs follow one of the proposed strategies.
We set $T=100$, $L=10$, and vary $N$ from $2$ to $10$ (recall that $N-1$ is the number of chaffs). The performance is averaged over $1000$ Monte Carlo runs.

\begin{figure}[tbh]
\small
\vspace{-.5em}
 \begin{minipage}{0.495\linewidth}
 \begin{center}
\includegraphics[width=1\linewidth]{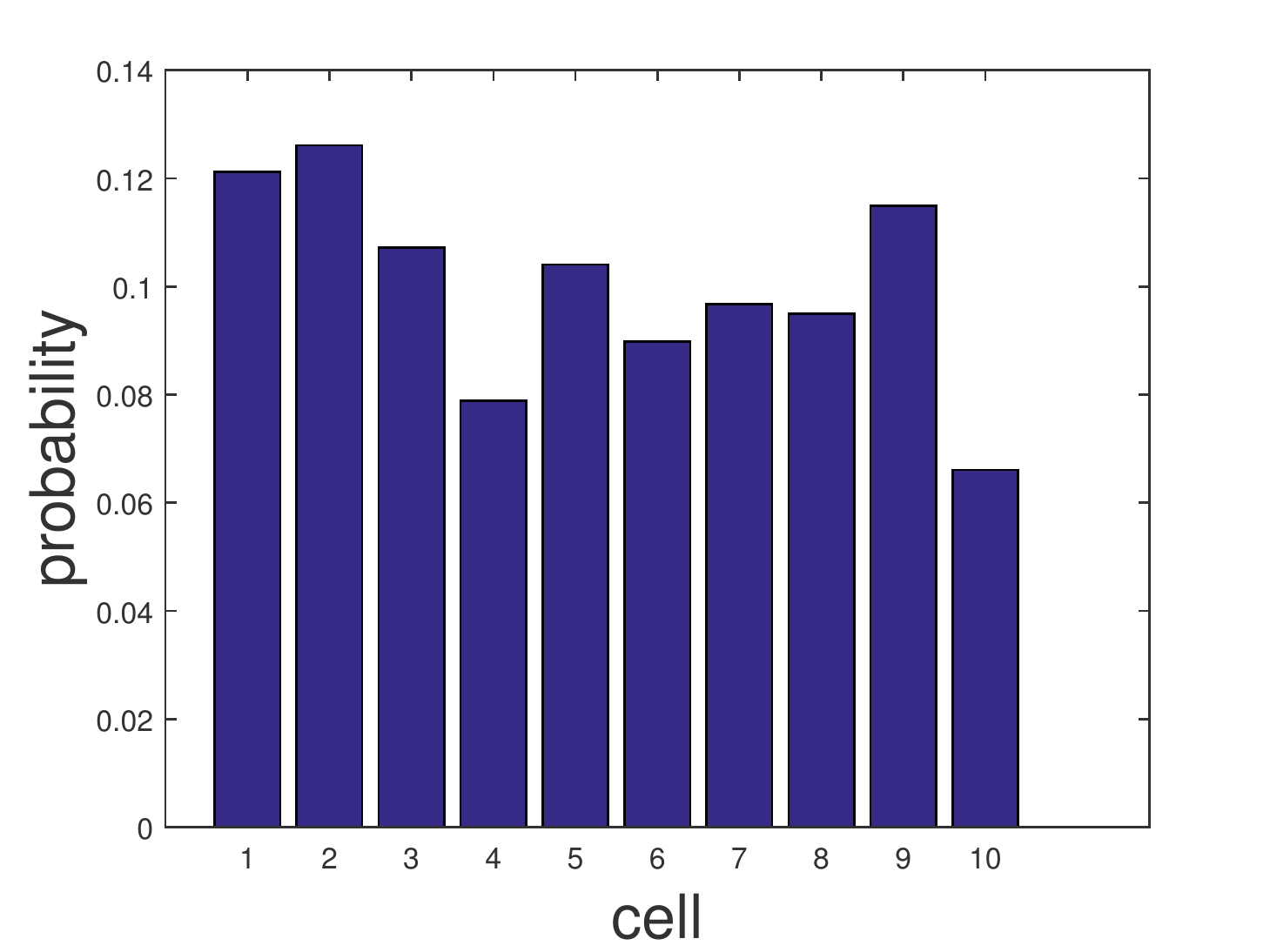}
\vspace{-1em}
\centerline{(a) non-skewed }
\end{center}
 \end{minipage}\hfill
 \begin{minipage}{0.495\linewidth}
 \begin{center}
\includegraphics[width=1\linewidth]{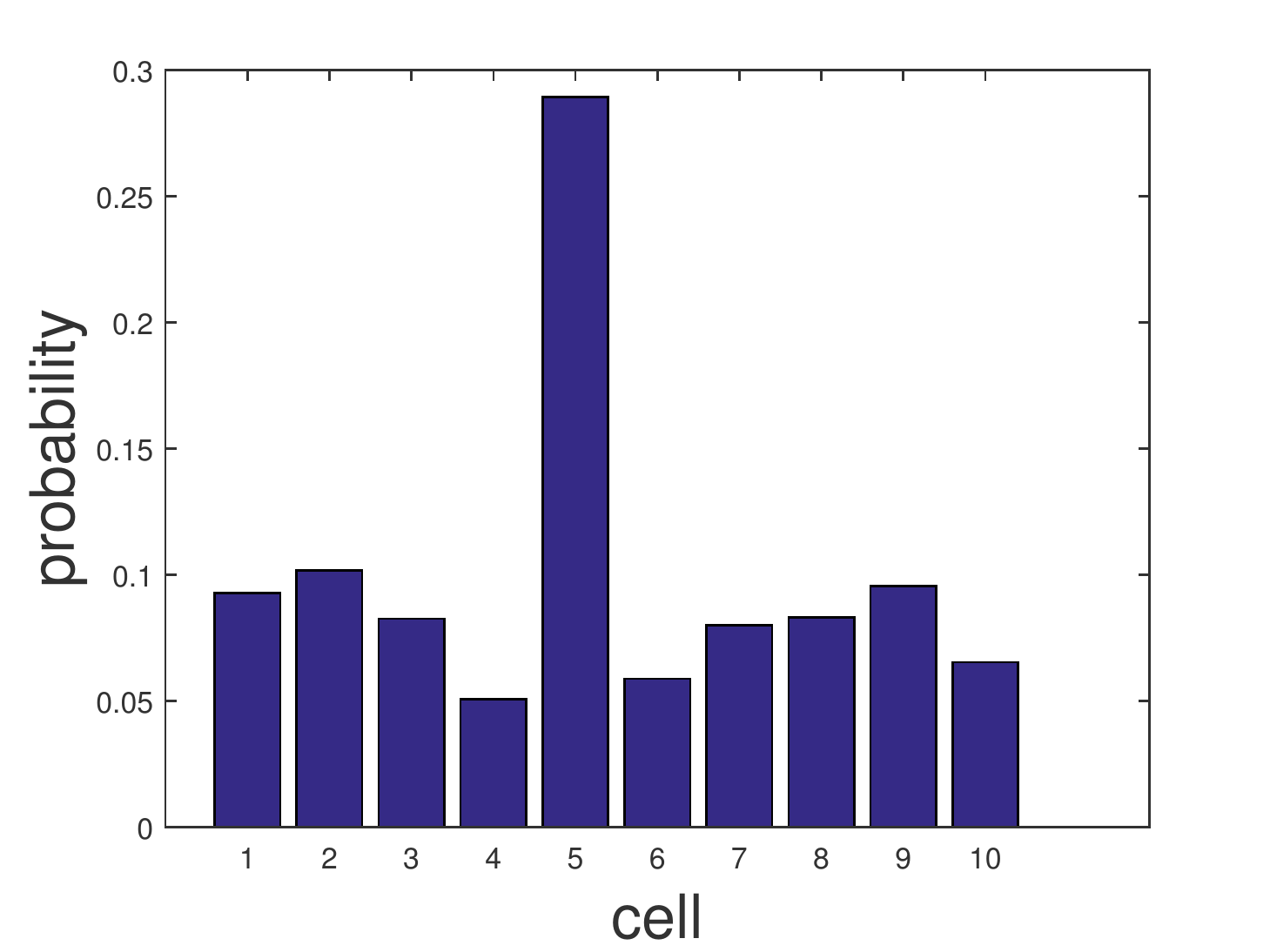}
\vspace{-1em}
\centerline{(b) spatially-skewed }
\end{center}
 \end{minipage}

 \vspace{1em}
 \begin{minipage}{0.495\linewidth}
 \begin{center}
\includegraphics[width=1\linewidth]{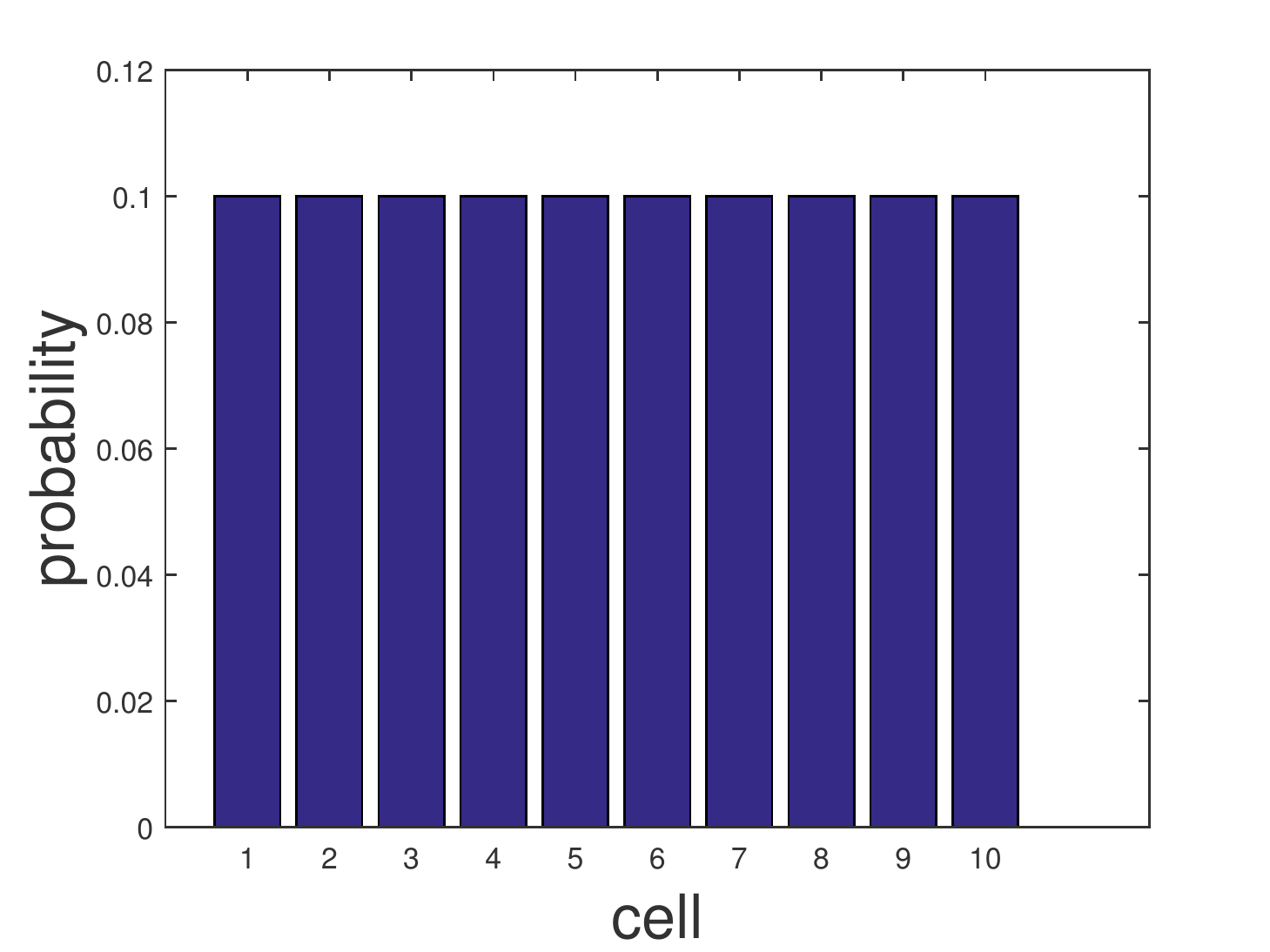}
\vspace{-1em}
\centerline{(c) temporally-skewed }
\end{center}
 \end{minipage}\hfill
 \begin{minipage}{0.495\linewidth}
 \begin{center}
\includegraphics[width=1\linewidth]{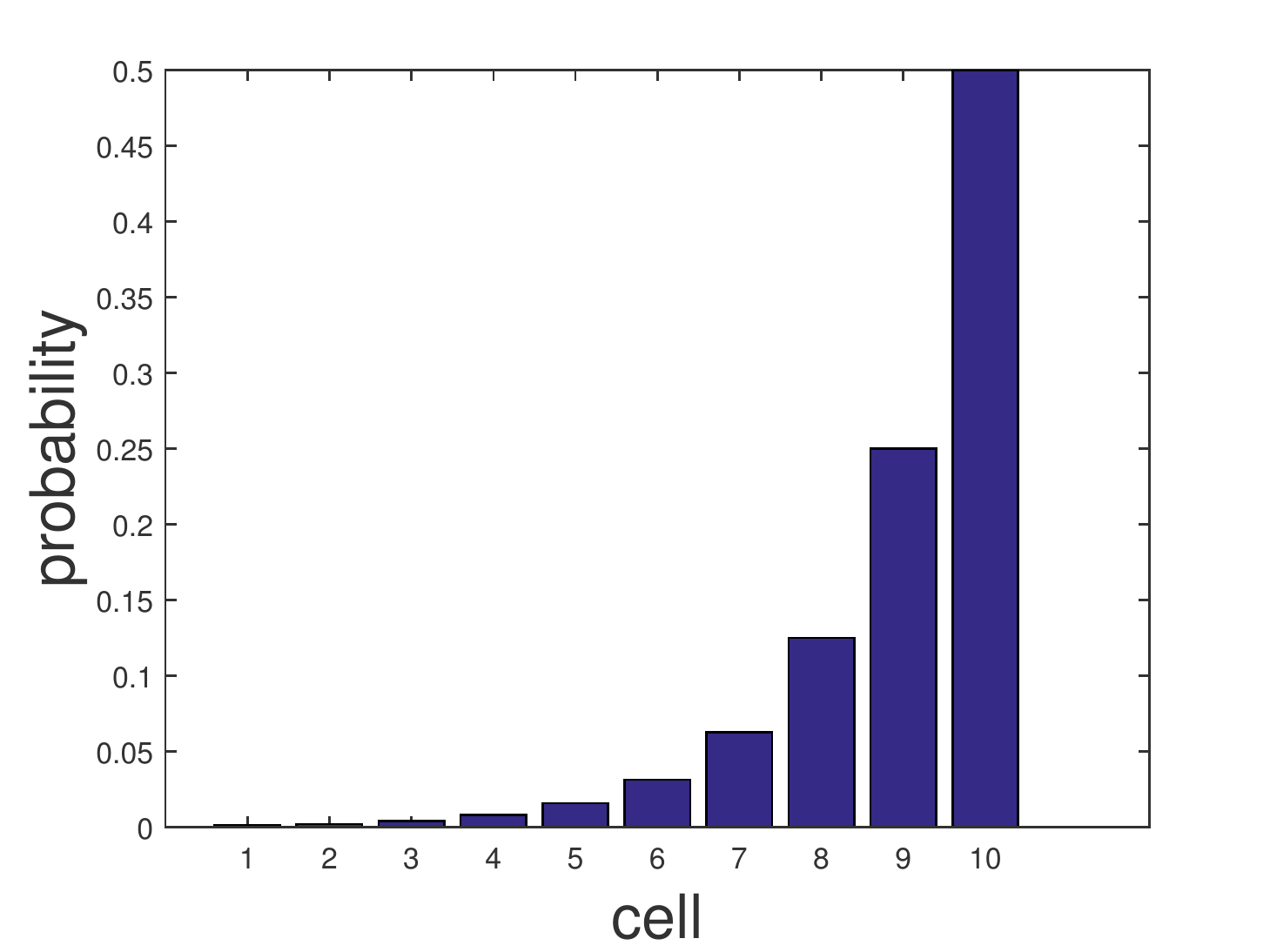}
\vspace{-1em}
\centerline{(d) spatially$\&$temporally-skewed }
\end{center}
 \end{minipage}
 \vspace{.0em}
 \caption{\small Steady-state distributions under various mobility models. } \label{fig:steady-state}
 \vspace{-.5em}
\end{figure}

We evaluate four different mobility models for the user: (a) \emph{neither} spatially \emph{nor} temporally skewed mobility, represented by a MC with randomly generated transition probabilities, (b) \emph{spatially}-skewed mobility, represented by a MC with a high probability of transiting into a certain cell\footnote{ This is generated by generating an $|\Lc|\times|\Lc|$ matrix of random values in $[0,1]$, setting the $j$-th ($j=5$) column to $2$, and normalizing each row. }, (c) \emph{temporally}-skewed mobility, represented by a random walk with a uniform steady-state distribution\footnote{This is generated by giving each cell probability $p$ of moving to the right, probability $q$ of moving to the left, and probability of $1-p-q$ of staying ($p=0.5$, $q=0.25$), and then wrapping transitions beyond the boundaries.}, and (d) \emph{both} spatially \emph{and} temporally skewed mobility, represented by a random walk with a non-uniform steady-state distribution\footnote{This is a variation of model (c) without wrapping at the boundaries. In models (c--d), we allow transitions between nonadjacent cells with $\epsilon$ probability ($\epsilon=10^{-5}$).}.
Fig.~\ref{fig:steady-state} gives the steady-state distribution under each model; its deviation from the uniform distribution measures the spatial skewness. To measure temporal skewness, we evaluate the average \emph{Kullback-Leibler (KL) distance}\footnote{The KL distance quantifies the difference between two probability distributions \cite{cover2012elements}.} between different rows of the transition matrix (the larger, the more skewed). The distances for models (a--d) are $0.44$, $0.34$, $8.18$, and $8.48$, respectively.

\subsubsection{Performance under Basic Eavesdropper}

\begin{figure}[tbh]
\small
\vspace{-.5em}
 \begin{minipage}{0.495\linewidth}
 \begin{center}
\includegraphics[width=1\linewidth]{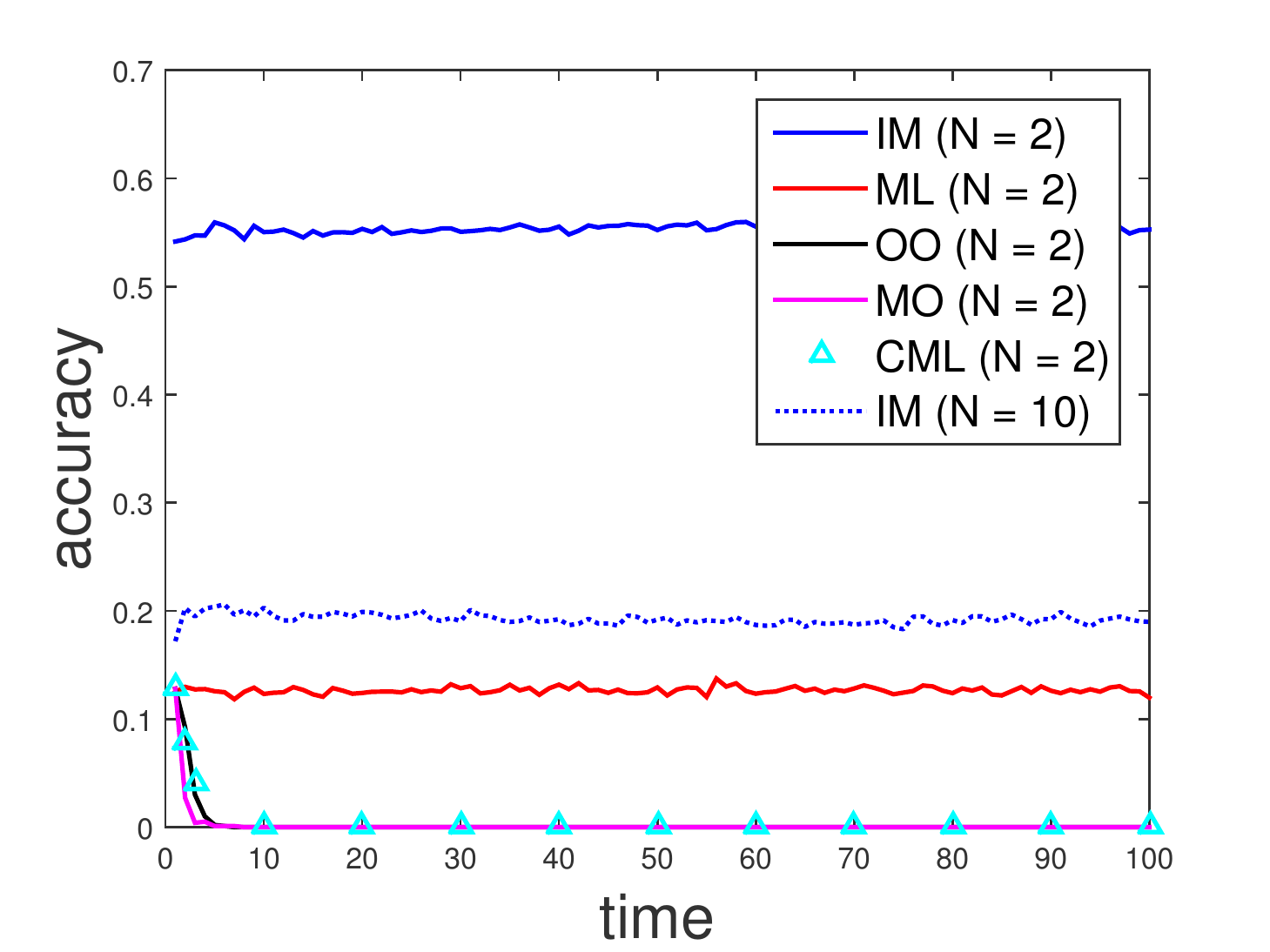}
\vspace{-1em}
\centerline{(a) non-skewed }
\end{center}
 \end{minipage}\hfill
 \begin{minipage}{0.495\linewidth}
 \begin{center}
\includegraphics[width=1\linewidth]{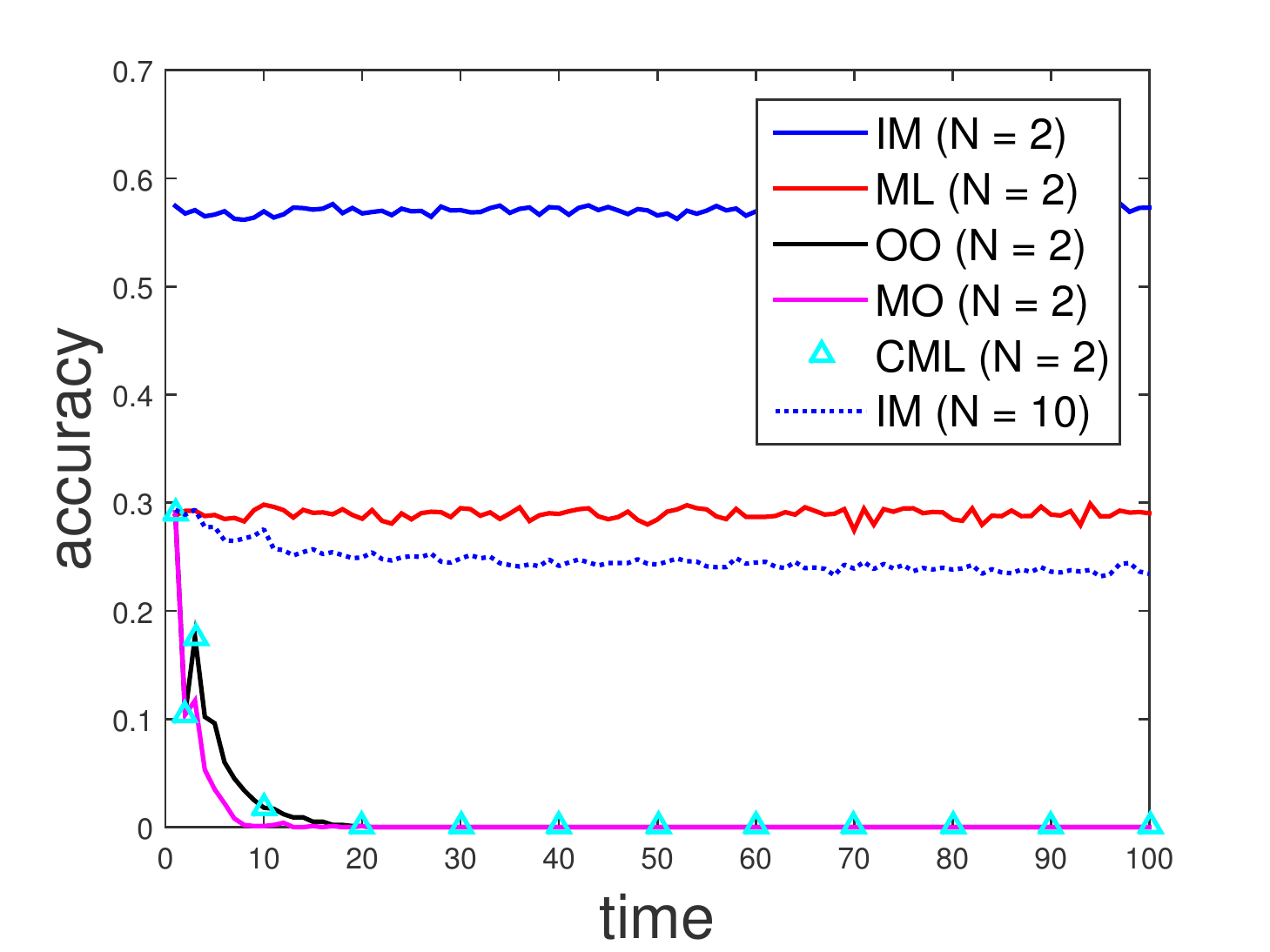}
\vspace{-1em}
\centerline{(b) spatially-skewed }
\end{center}
 \end{minipage}

 \vspace{1em}
 \begin{minipage}{0.495\linewidth}
 \begin{center}
\includegraphics[width=1\linewidth]{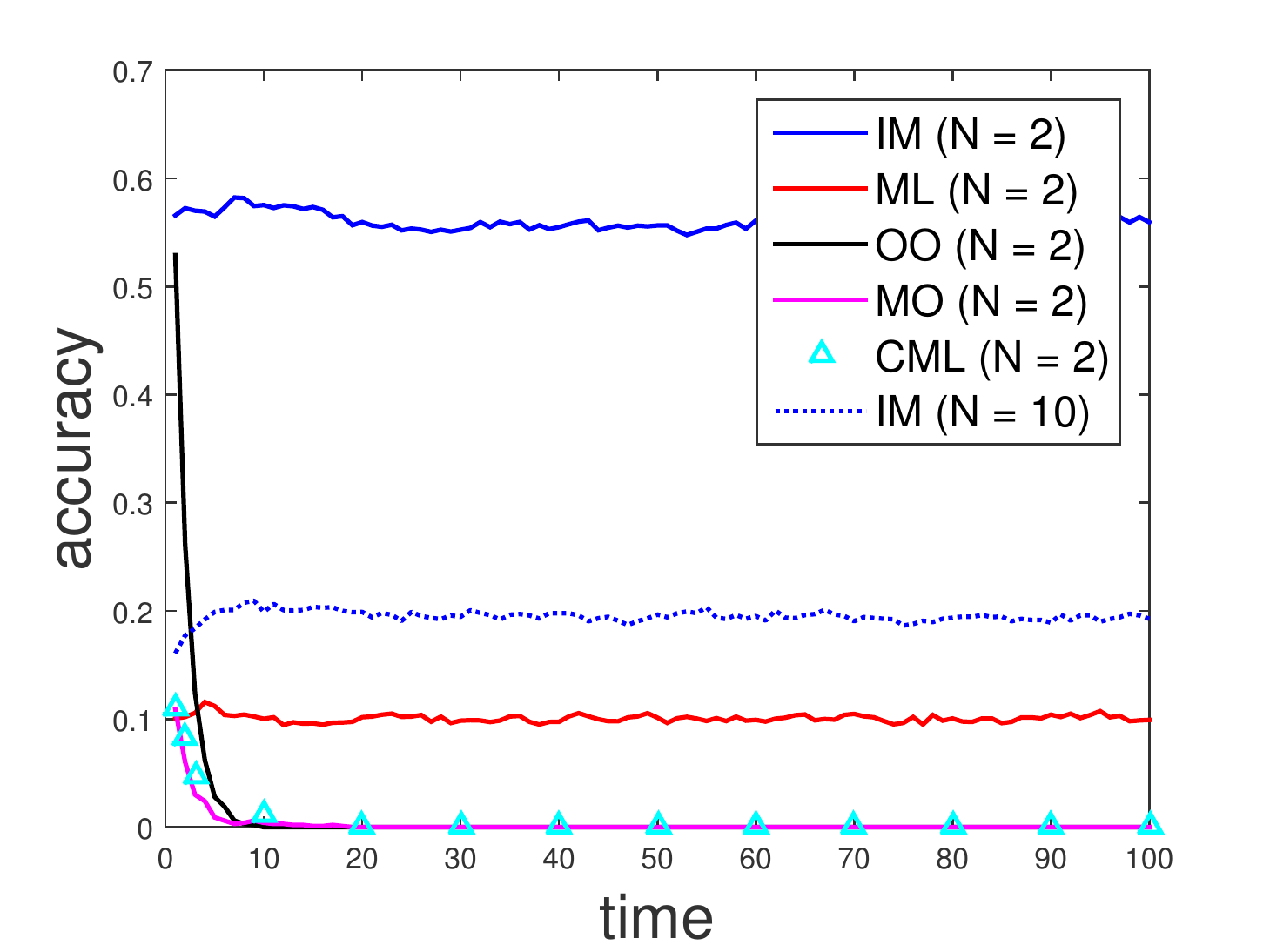}
\vspace{-1em}
\centerline{(c) temporally-skewed }
\end{center}
 \end{minipage}\hfill
 \begin{minipage}{0.495\linewidth}
 \begin{center}
\includegraphics[width=1\linewidth]{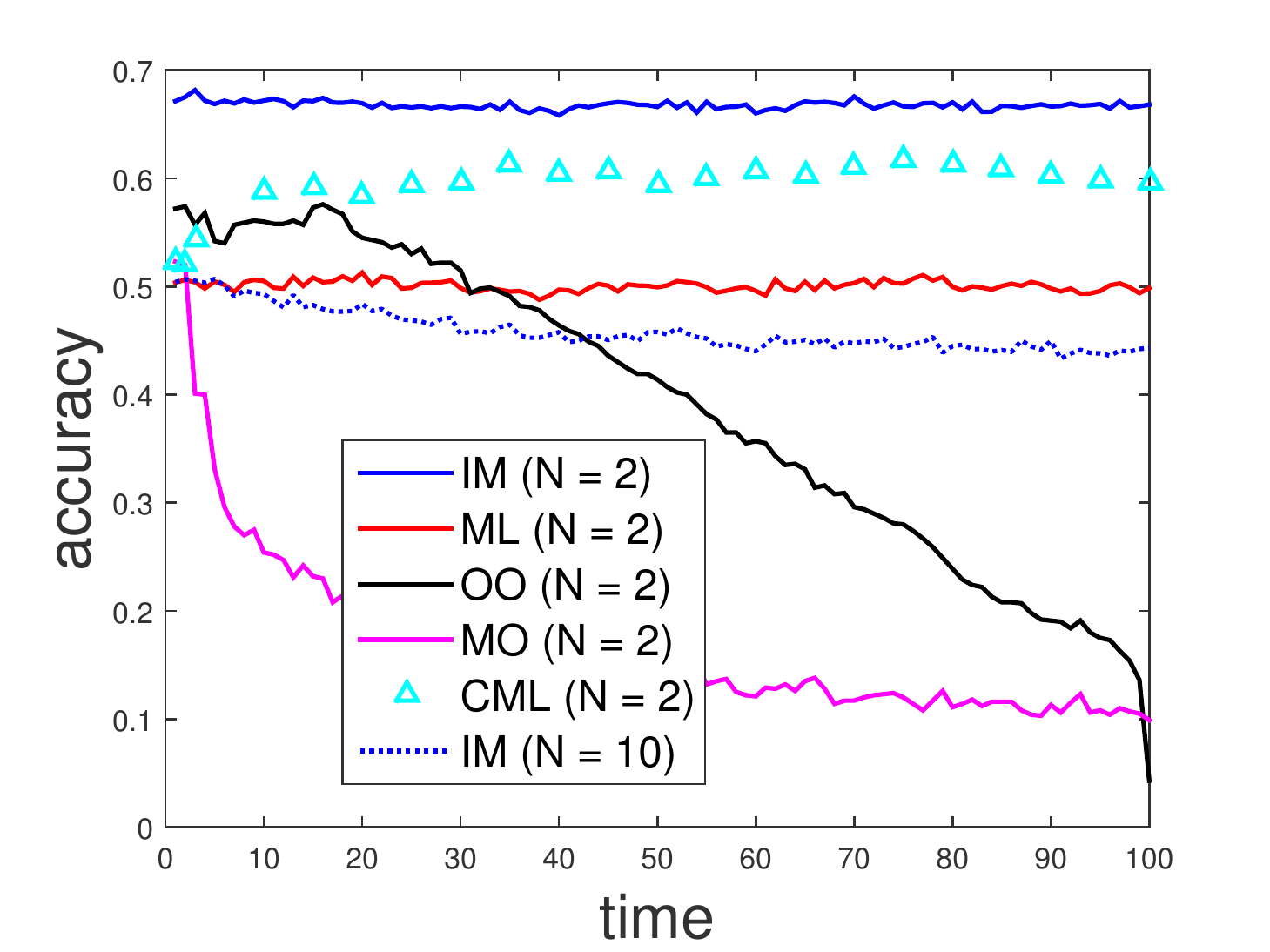}
\vspace{-1em}
\centerline{(d) spatially$\&$temporally-skewed }
\end{center}
 \end{minipage}
 \vspace{.0em}
 \caption{\small Tracking accuracy of basic eavesdropper. } \label{fig:accuracy_basic}
 \vspace{-.5em}
\end{figure}

\begin{figure}[tbh]
\small
\vspace{-.5em}
 \begin{minipage}{0.495\linewidth}
 \begin{center}
\includegraphics[width=1\linewidth]{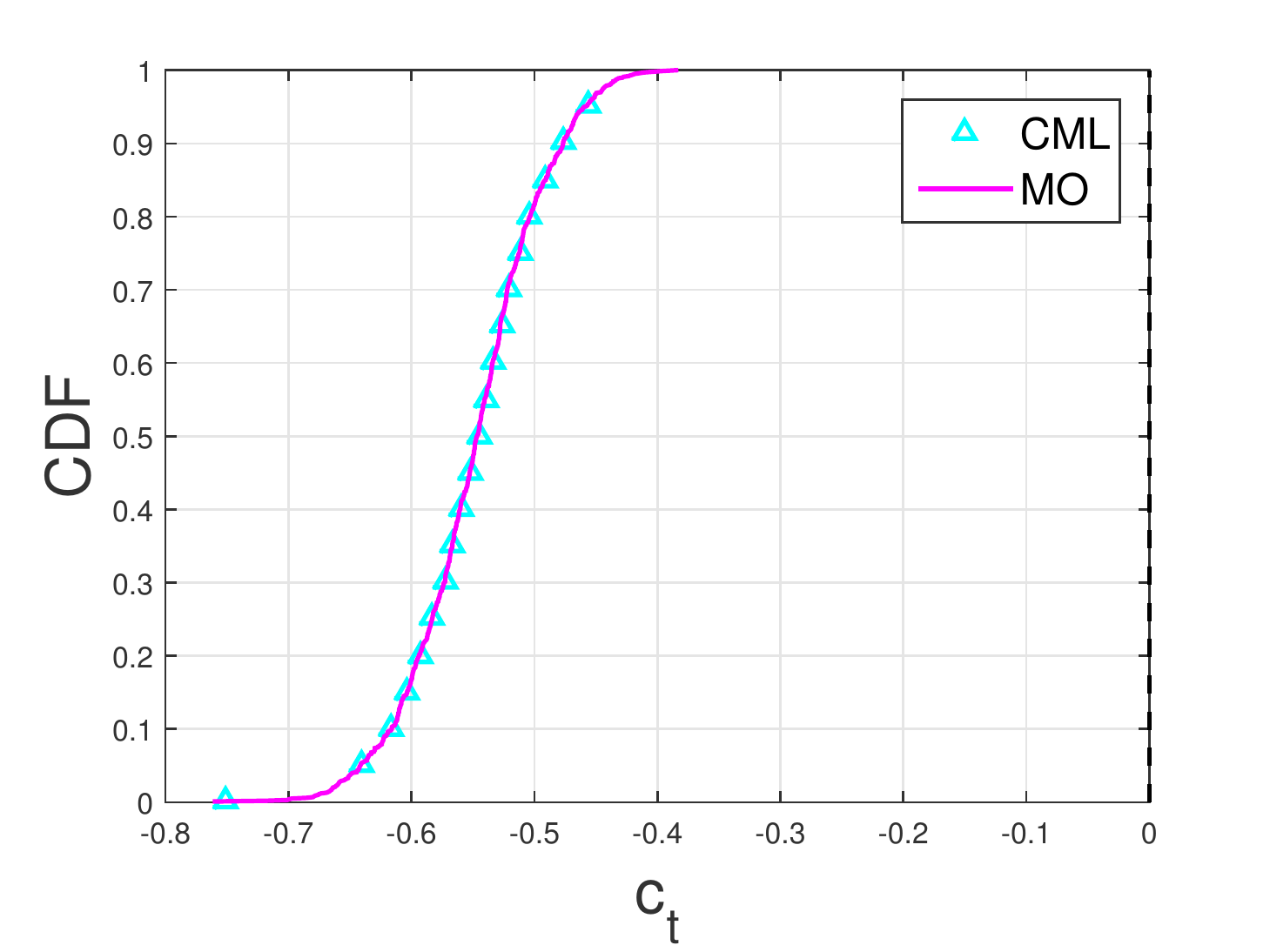}
\vspace{-1em}
\centerline{(a) non-skewed }
\end{center}
 \end{minipage}\hfill
 \begin{minipage}{0.495\linewidth}
 \begin{center}
\includegraphics[width=1\linewidth]{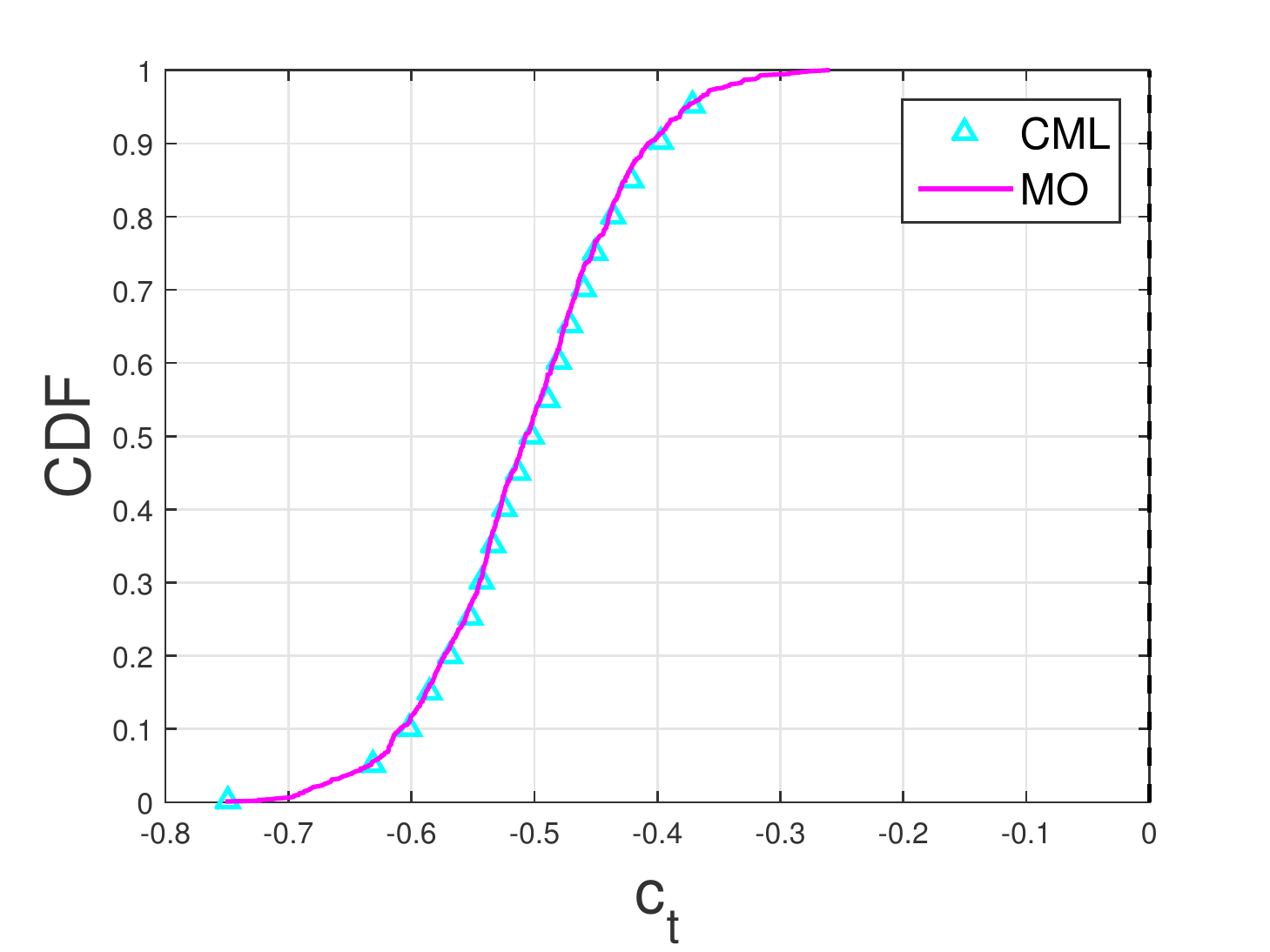}
\vspace{-1em}
\centerline{(b) spatially-skewed }
\end{center}
 \end{minipage}

 \vspace{1em}
 \begin{minipage}{0.495\linewidth}
 \begin{center}
\includegraphics[width=1\linewidth]{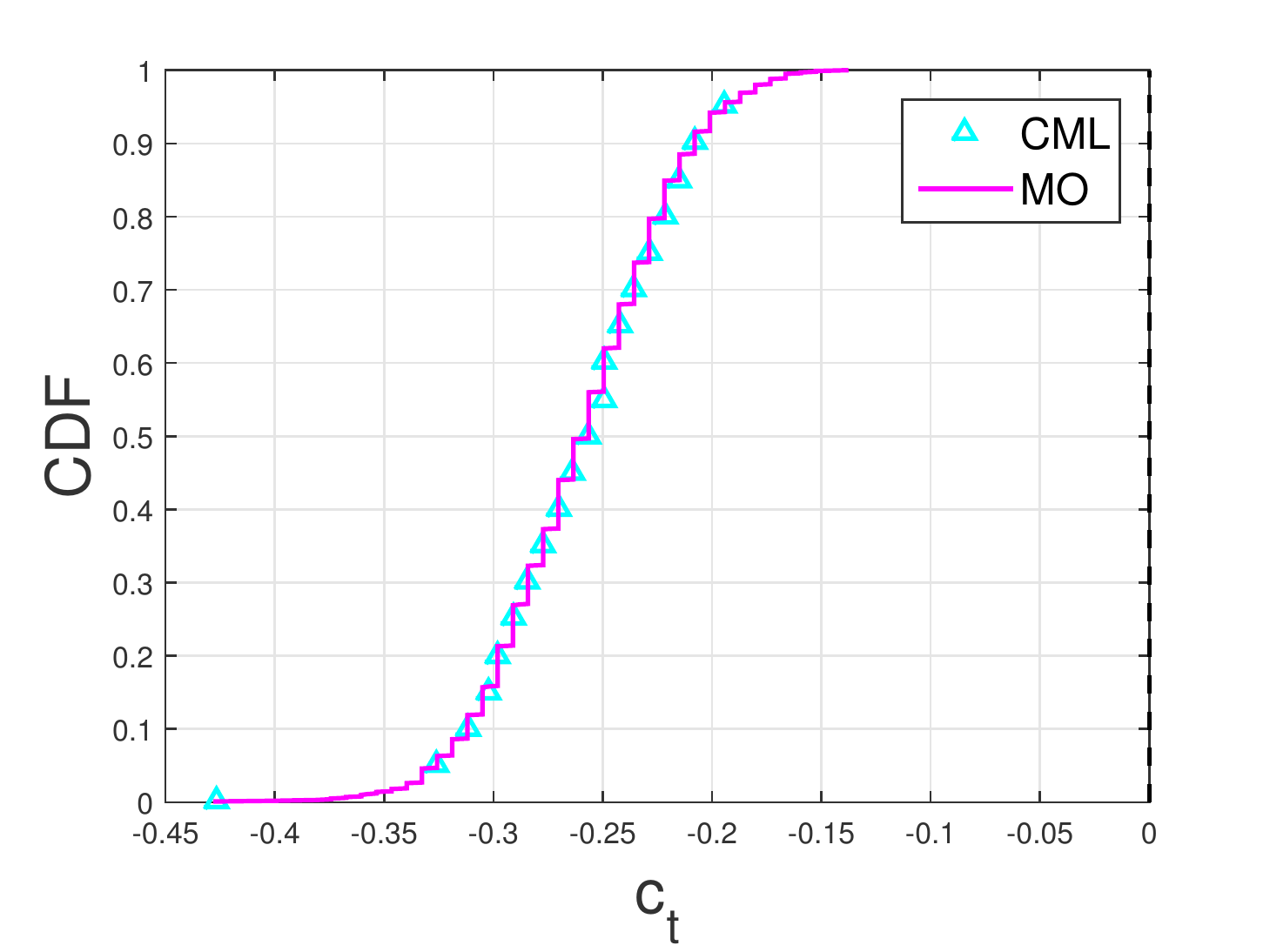}
\vspace{-1em}
\centerline{(c) temporally-skewed }
\end{center}
 \end{minipage}\hfill
 \begin{minipage}{0.495\linewidth}
 \begin{center}
\includegraphics[width=1\linewidth]{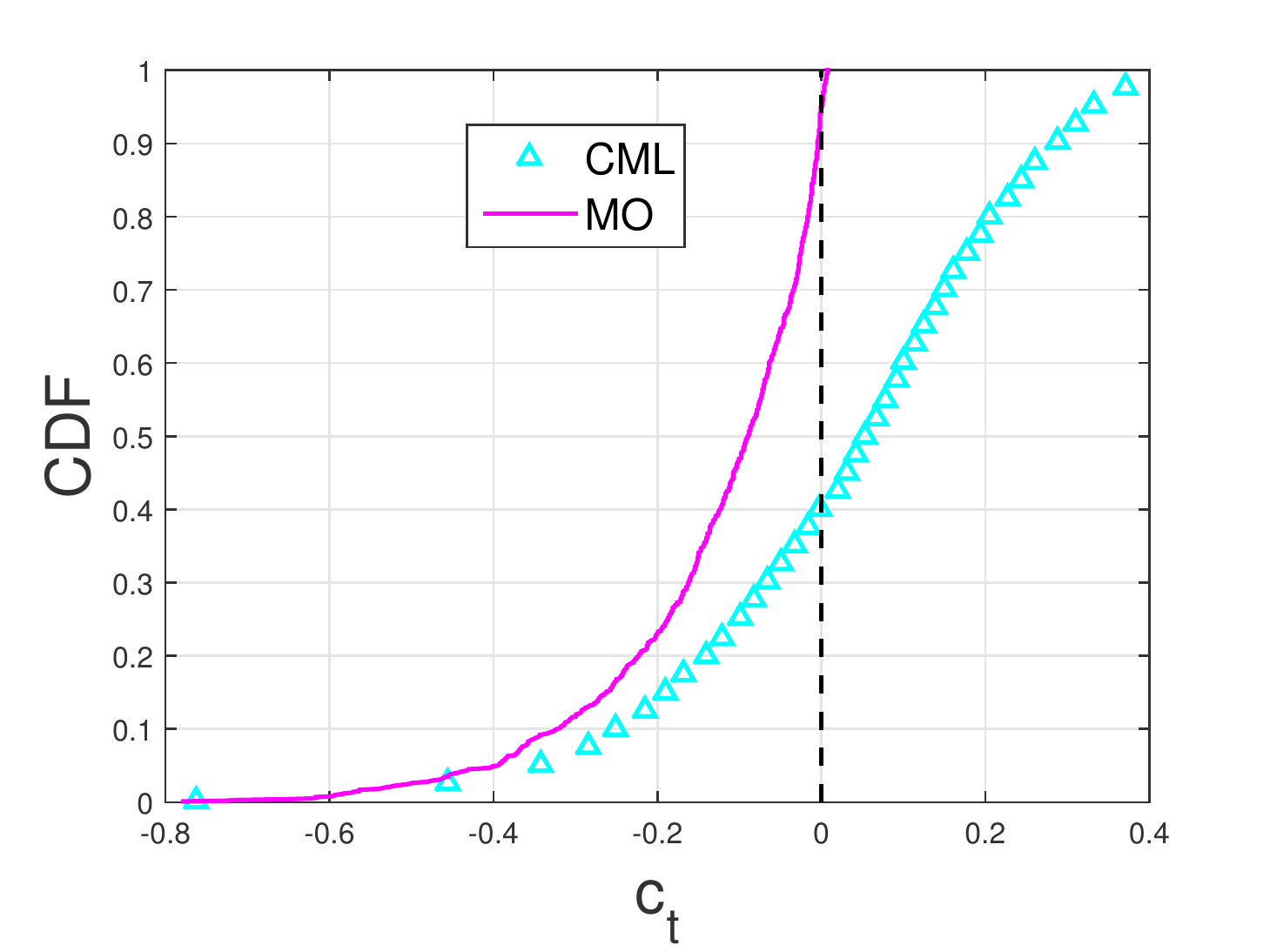}
\vspace{-1em}
\centerline{(d) spatially$\&$temporally-skewed }
\end{center}
 \end{minipage}
 \vspace{.0em}
 \caption{\small Distribution of $c_t$ (defined in (\ref{eq:c_1}, \ref{eq:c_t})). } \label{fig:ct_basic}
 \vspace{-.5em}
\end{figure}

We first evaluate the performance of a basic eavesdropper using an ML detector (\ref{eq:ML - single-user}); see Fig.~\ref{fig:accuracy_basic}. We see that: (i) while IM and ML always lead to non-zero tracking accuracy, OO and MO can drive the tracking accuracy to zero over a sufficiently long time; (ii) the more skewed the mobility model (i.e., the more predictable the user movements), the higher the tracking accuracy; (iii) while the deterministic strategies (ML, OO, MO) cannot benefit from using more chaffs, the IM strategy can use more chaffs to lower the tracking accuracy. We further simulate the auxiliary strategy CML in Section~\ref{subsec:Accuracy - OO Strategy} and verify our analysis that the accuracy under CML/MO decays exponentially if $\mbbE[c_t]<0$; see Fig.~\ref{fig:ct_basic}. 
Note that OO is designed to be optimal over $T$ slots, and is not necessarily optimal for each $t<T$. Indeed, MO achieves a lower accuracy at $t<T$ in  Fig.~\ref{fig:accuracy_basic}~(d).

\subsubsection{Performance under Advanced Eavesdropper}

\begin{figure}[tbh]
\small
\vspace{-.5em}
 \begin{minipage}{0.495\linewidth}
 \begin{center}
\includegraphics[width=1\linewidth]{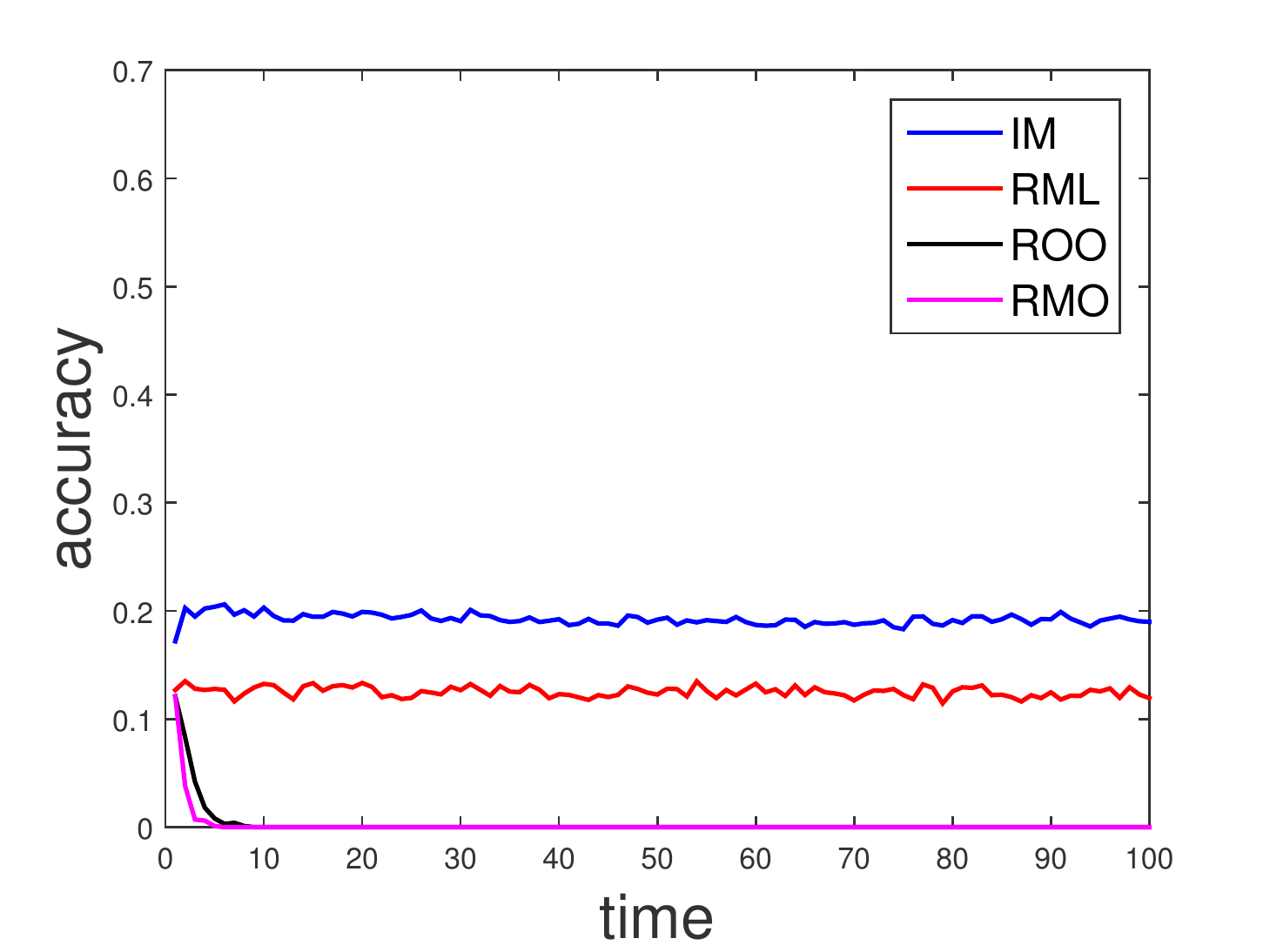}
\vspace{-1em}
\centerline{(a) non-skewed }
\end{center}
 \end{minipage}\hfill
 \begin{minipage}{0.495\linewidth}
 \begin{center}
\includegraphics[width=1\linewidth]{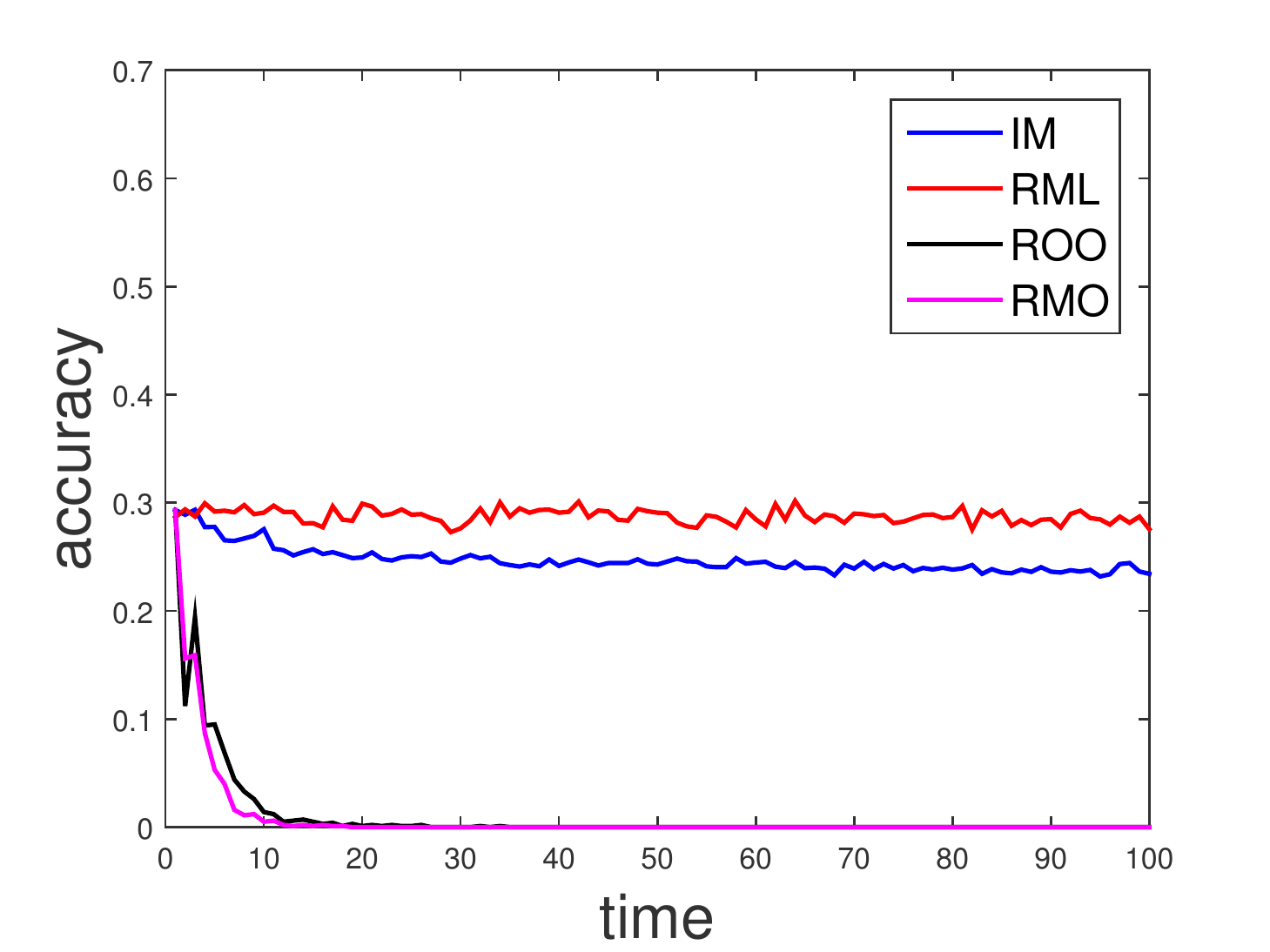}
\vspace{-1em}
\centerline{(b) spatially-skewed }
\end{center}
 \end{minipage}

 \vspace{1em}
 \begin{minipage}{0.495\linewidth}
 \begin{center}
\includegraphics[width=1\linewidth]{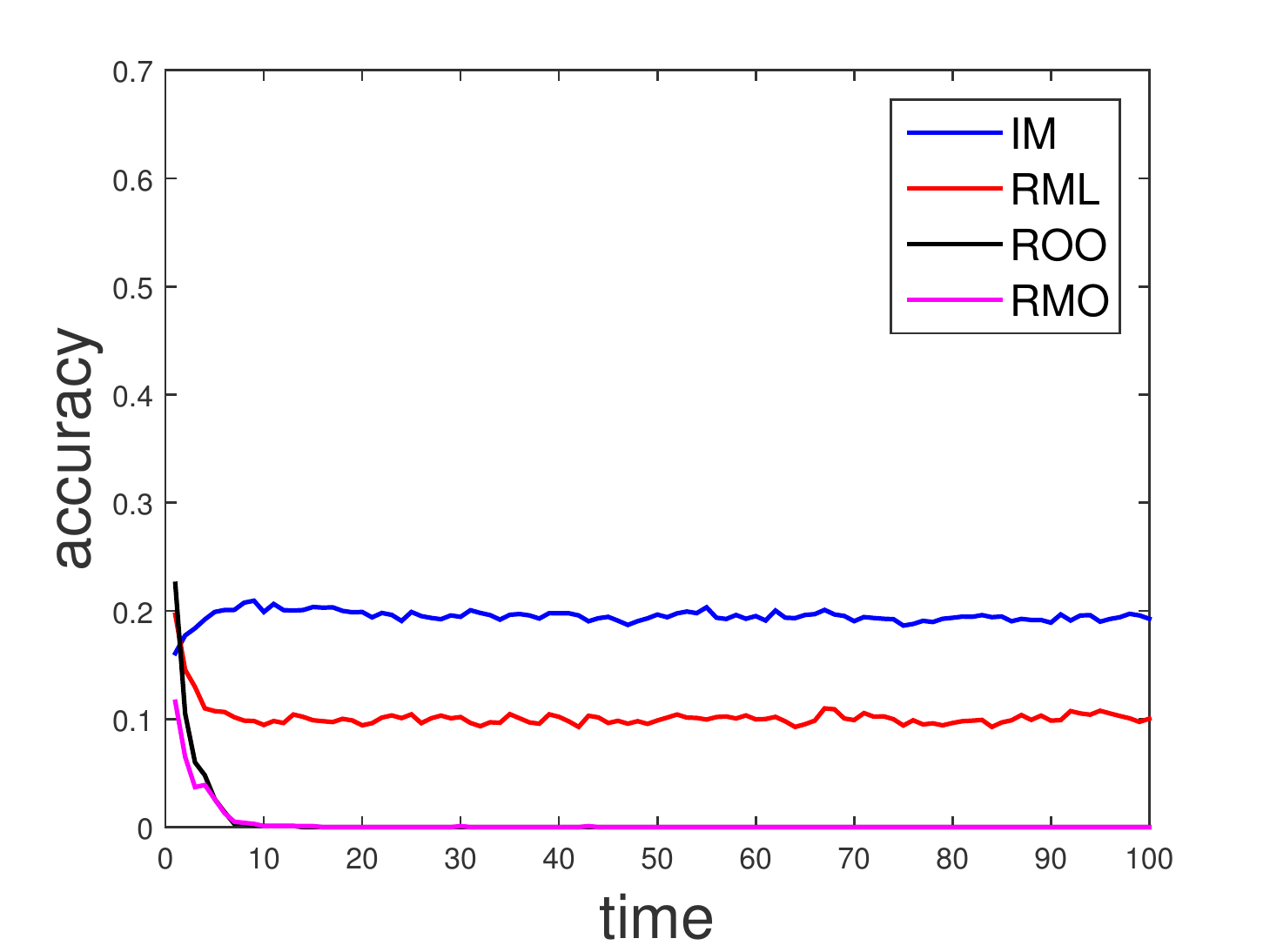}
\vspace{-1em}
\centerline{(c) temporally-skewed }
\end{center}
 \end{minipage}\hfill
 \begin{minipage}{0.495\linewidth}
 \begin{center}
\includegraphics[width=1\linewidth]{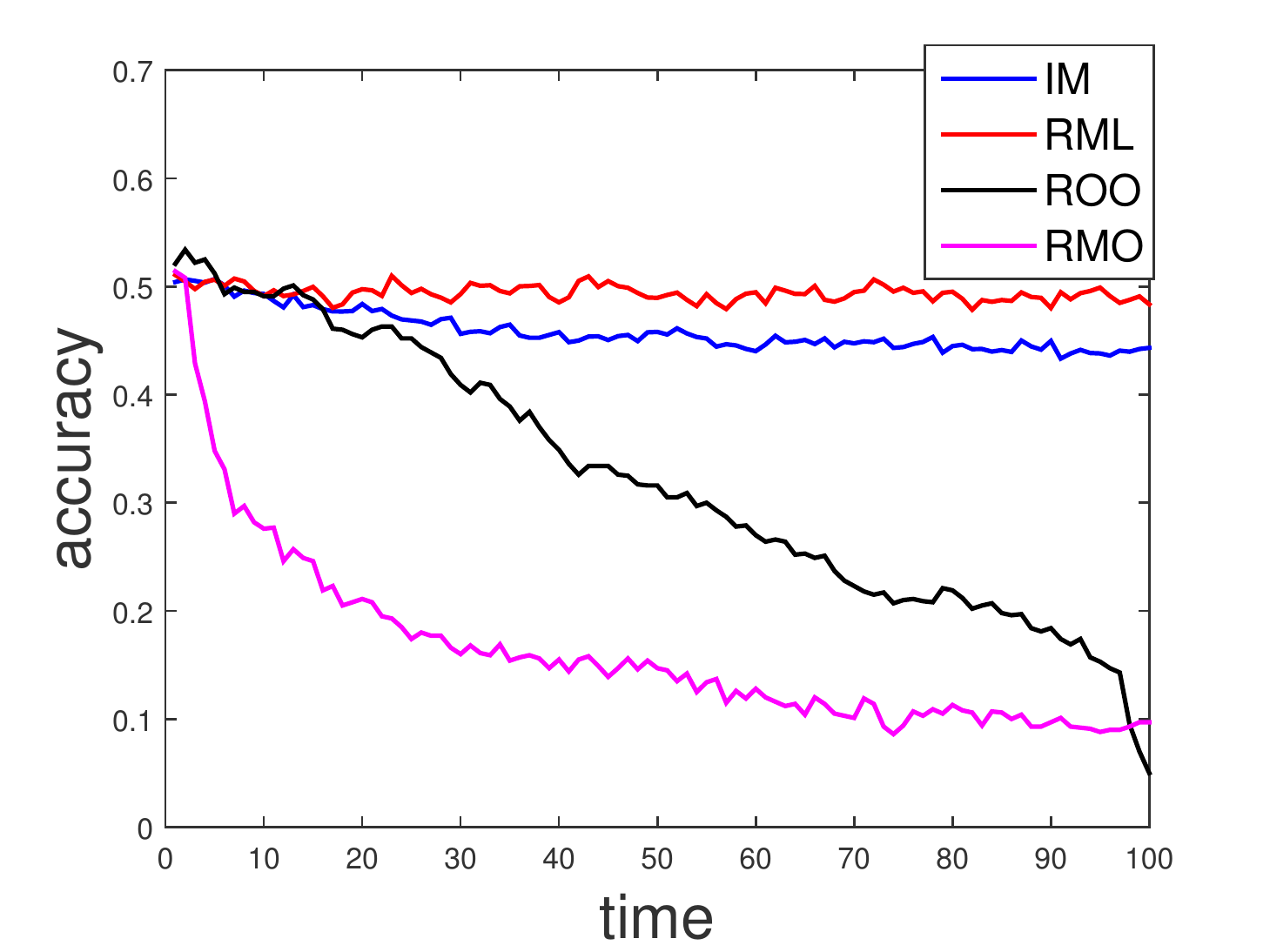}
\vspace{-1em}
\centerline{(d) spatially$\&$temporally-skewed }
\end{center}
 \end{minipage}
 \vspace{.0em}
 \caption{\small Tracking accuracy of advanced eavesdropper ($N=10$). } \label{fig:accuracy_advanced}
 \vspace{-.5em}
\end{figure}

We then evaluate the performance of an advanced eavesdropper aware of the strategy. Assume that the advanced eavesdropper first filters out trajectories matching the chaff's trajectory and then performs ML detection on the remaining trajectories. As expected, the deterministic strategies (ML, OO, MO) are ineffective against such an eavesdropper (not shown).
We thus focus on the IM strategy and the robust strategies (RML, ROO, RMO) in Section~\ref{subsec:Defense against Adanced Eavesdropper}; see Fig.~\ref{fig:accuracy_advanced}.
We see that by slightly perturbing the chaff's trajectory, the robust strategies not only prevent the chaffs from being recognized by the eavesdropper but also mimic the performance of their deterministic counterparts under a basic eavesdropper.

\subsection{Trace-driven Simulations}\label{subsec:Trace-driven Simulations}

\subsubsection{Dataset}

\begin{figure}[tbh]
\small
\vspace{-.5em}
 \begin{minipage}{0.495\linewidth}
 \begin{center}
\includegraphics[width=0.975\linewidth]{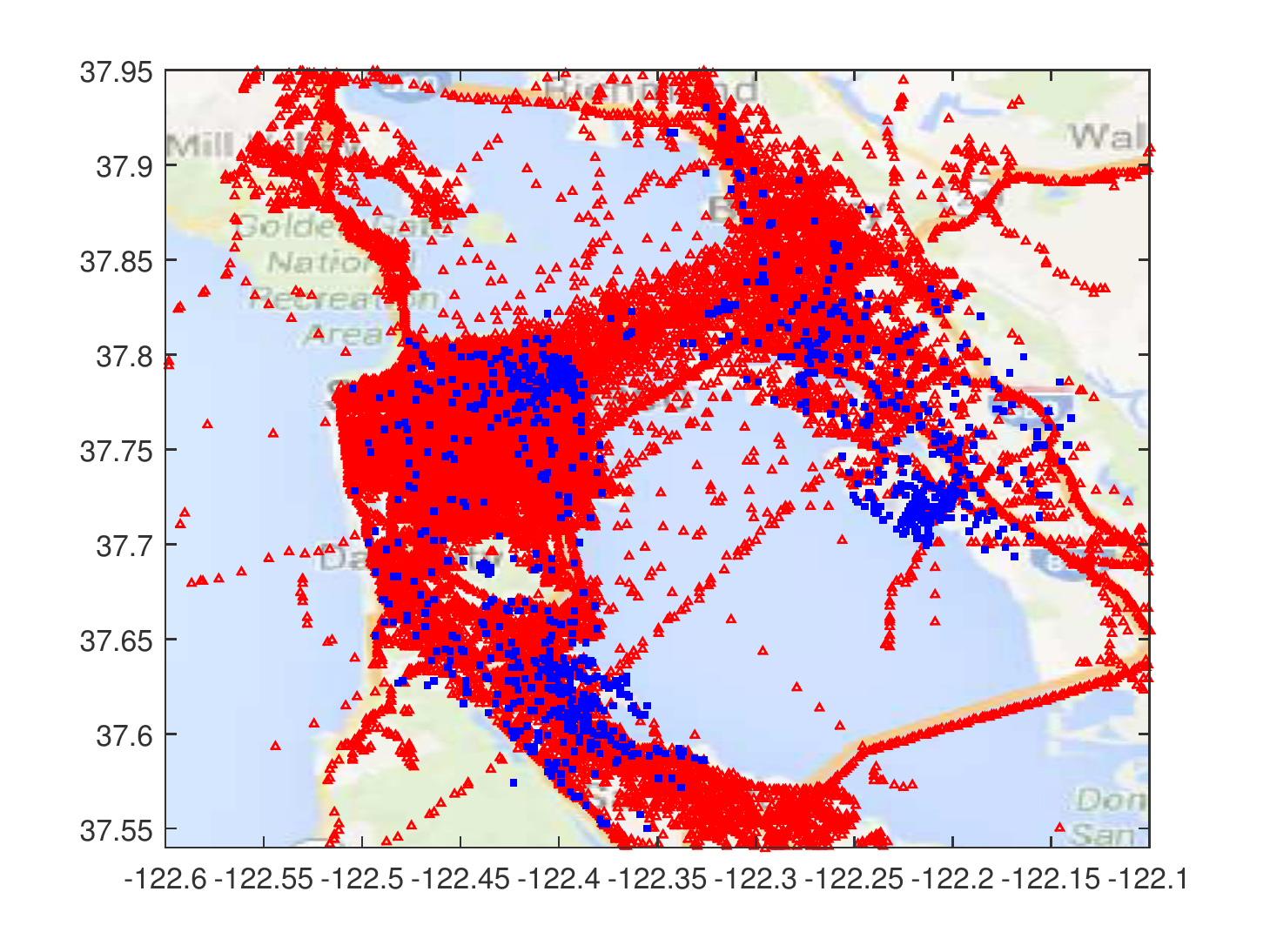}
\vspace{-1em}
\centerline{(a) \textcolor{red}{$\blacktriangle$}: node; \textcolor{blue}{$\blacksquare$}: cell tower. }
\end{center}
 \end{minipage}\hfill
 \begin{minipage}{0.495\linewidth}
 \begin{center}
\includegraphics[width=1\linewidth]{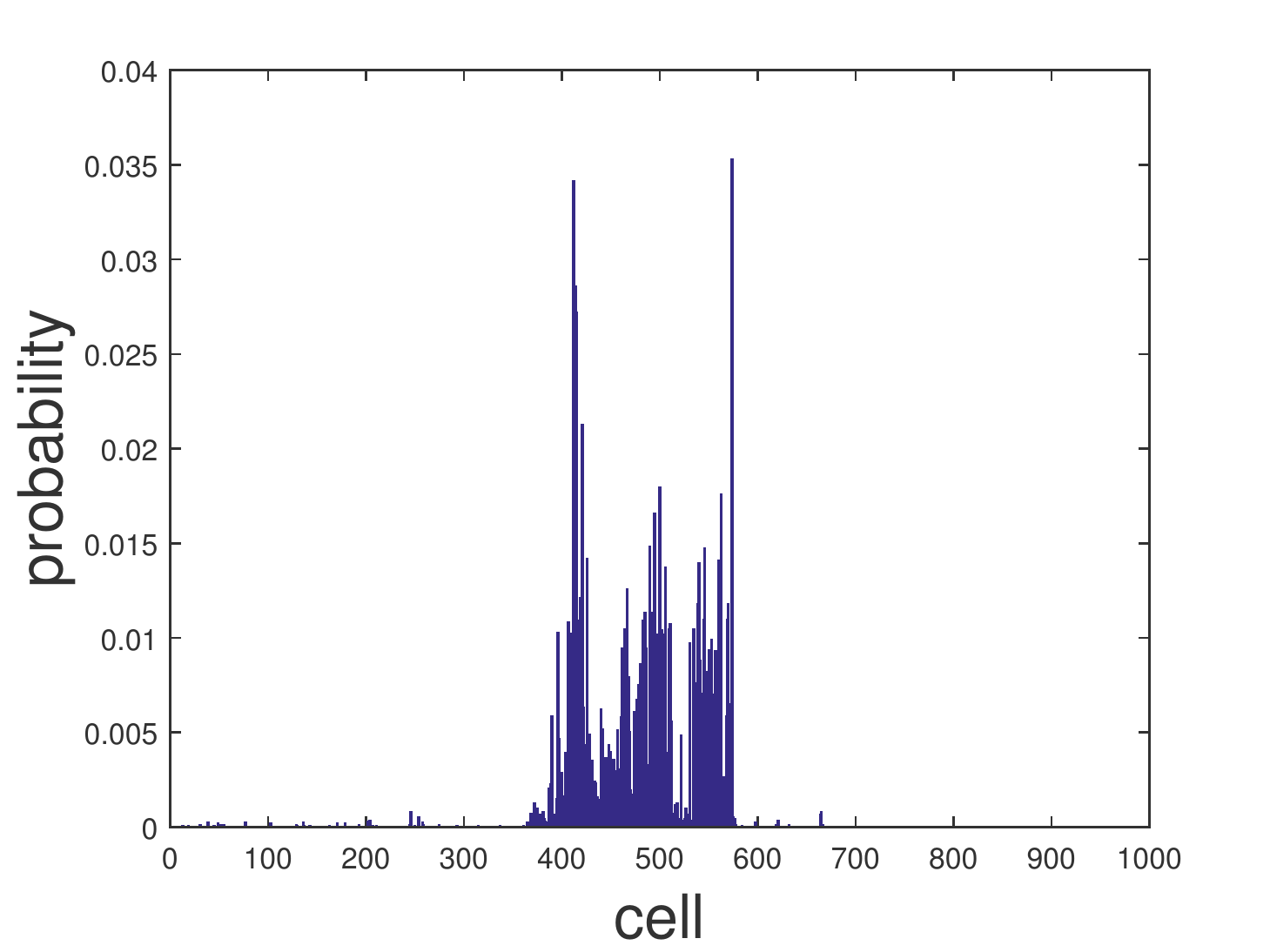}
\vspace{-1em}
\centerline{(b) steady-state distribution }
\end{center}
 \end{minipage}
 \vspace{.0em}
 \caption{\small Cell layout and steady-state distribution. } \label{fig:cell_setting_realtower}
 \vspace{-.5em}
\end{figure}

We further evaluate our solutions on real mobility traces. We use the taxi cab traces from \cite{SFTaxidata}, from which we extract the traces of $174$ nodes over a $100$-minute period with location updates every minute\footnote{The traces have irregular update intervals. We filter out inactive nodes (no update for $5$ minutes) and regulate the intervals through linear interpolation. }. We quantize the node locations into $959$ \emph{Voronoi cells} based on cell tower locations obtained from \texttt{http://www.antennasearch.com} (ignoring towers within $100$ meters of others); see Fig.~\ref{fig:cell_setting_realtower}~(a).
Modeling the $174$ traces as trajectories generated independently from the same MC, we compute the empirical transition matrix and the empirical steady-state distribution (Fig.~\ref{fig:cell_setting_realtower}~(b)).
Clearly, this mobility model is spatially-skewed; we have verified that it is also temporally-skewed. 

\subsubsection{Performance under Basic Eavesdropper}

\begin{figure}[tbh]
\small
\vspace{-.5em}
 \begin{minipage}{0.495\linewidth}
 \begin{center}
\includegraphics[width=1\linewidth]{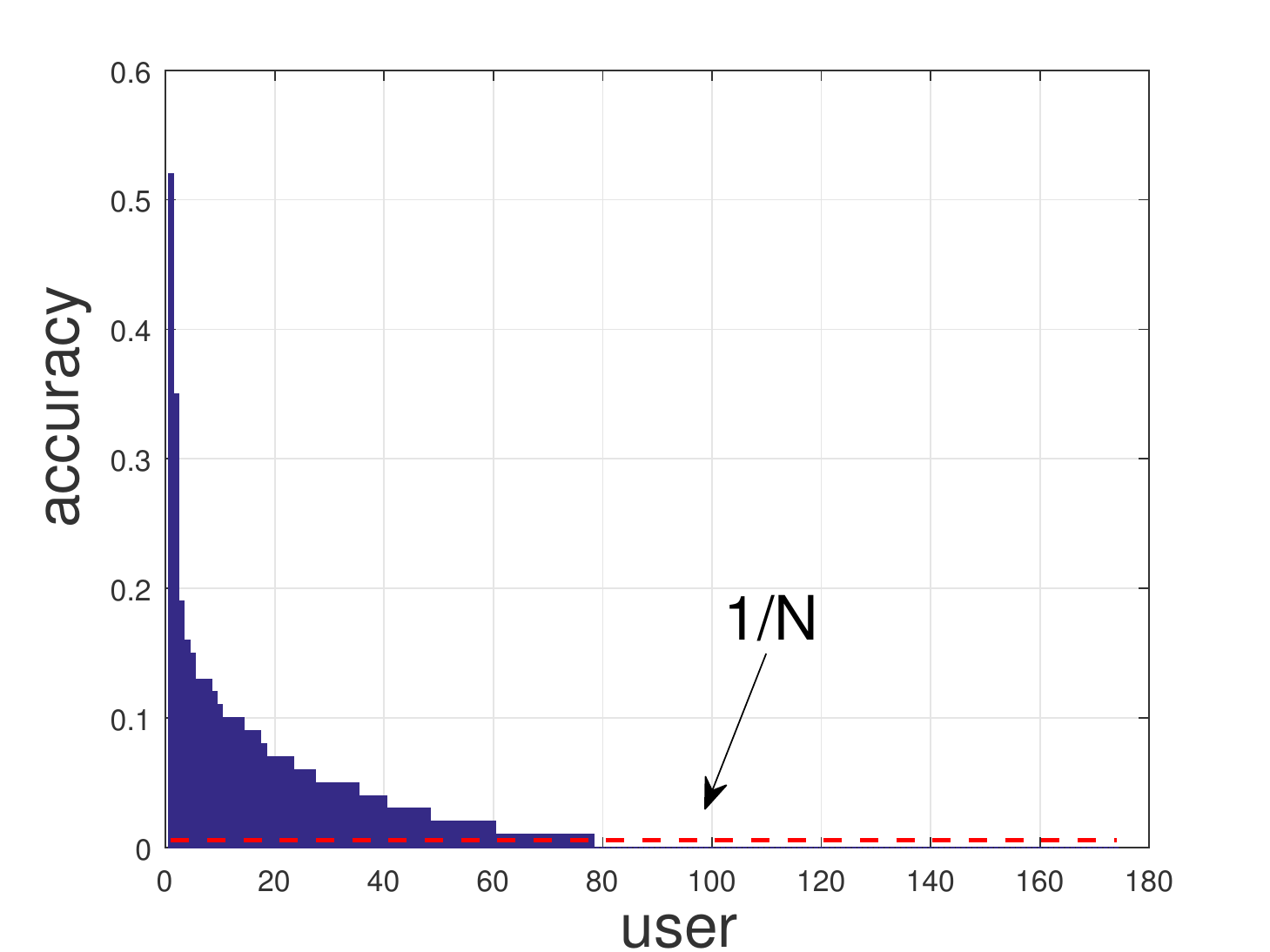}
\vspace{-1em}
\centerline{(a) no chaff }
\end{center}
 \end{minipage}\hfill
 \begin{minipage}{0.495\linewidth}
 \begin{center}
\includegraphics[width=1\linewidth]{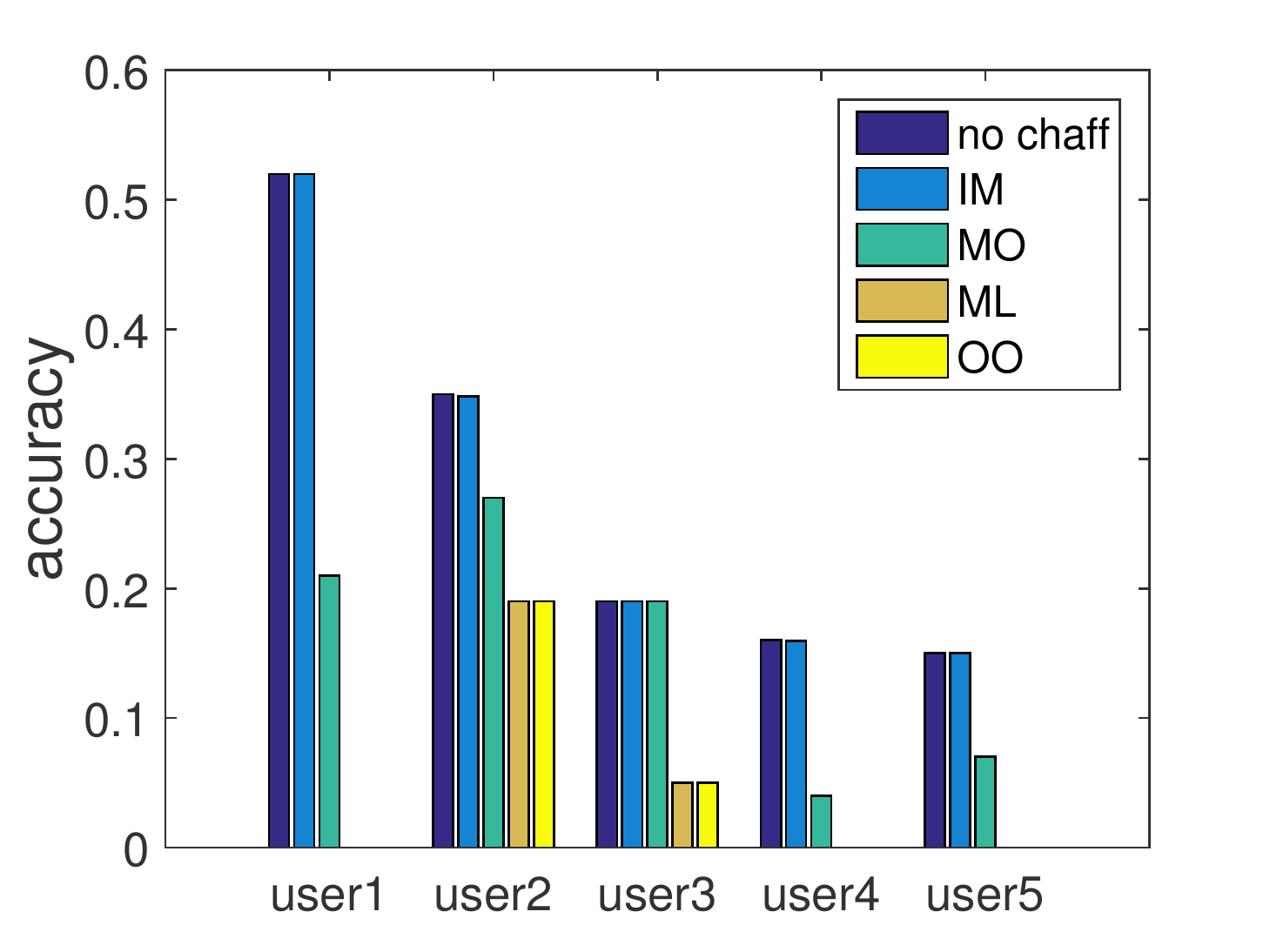}
\vspace{-1em}
\centerline{(b) a single chaff }
\end{center}
 \end{minipage}
 \vspace{.0em}
 \caption{\small Accuracy of basic eavesdropper before/after adding chaff (missing bars denote zero accuracy). } \label{fig:accuracy_basic_trace}
 \vspace{-.5em}
\end{figure}

We first evaluate a basic eavesdropper running ML detection. When there is no chaff as shown in Fig.~\ref{fig:accuracy_basic_trace}~(a), the eavesdropper can track a subset of users with much higher accuracy than the baseline of $1/N\approx 0.6\%$. For example, user $1$ is tracked $52\%$ of the time, and users $2,\ldots,5$ are tracked at least $15\%$ of the time.
We then evaluate the accuracy in tracking the top-$K$ users ($K=5$) after adding a single chaff in Fig.~\ref{fig:accuracy_basic_trace}~(b). While IM cannot help these users, ML and OO can significantly lower the tracking accuracy.
Meanwhile, MO performs relatively poorly for these users, because their trajectories jointly dominate the myopic trajectory in likelihood for $55\%$ of the time, during which MO cannot alter the decision of the detector. ML and OO avoid this problem by not moving to the ML location at the beginning.
Note that this limitation of MO can be overcome by using more sophisticated solvers to the MDP in Section~\ref{subsec:Online Strategy}; detailed evaluations are left to future work.

\subsubsection{Performance under Advanced Eavesdropper}

\begin{figure}[tbh]
\small
\vspace{-.5em}
 \begin{center}
\includegraphics[width=0.5\linewidth]{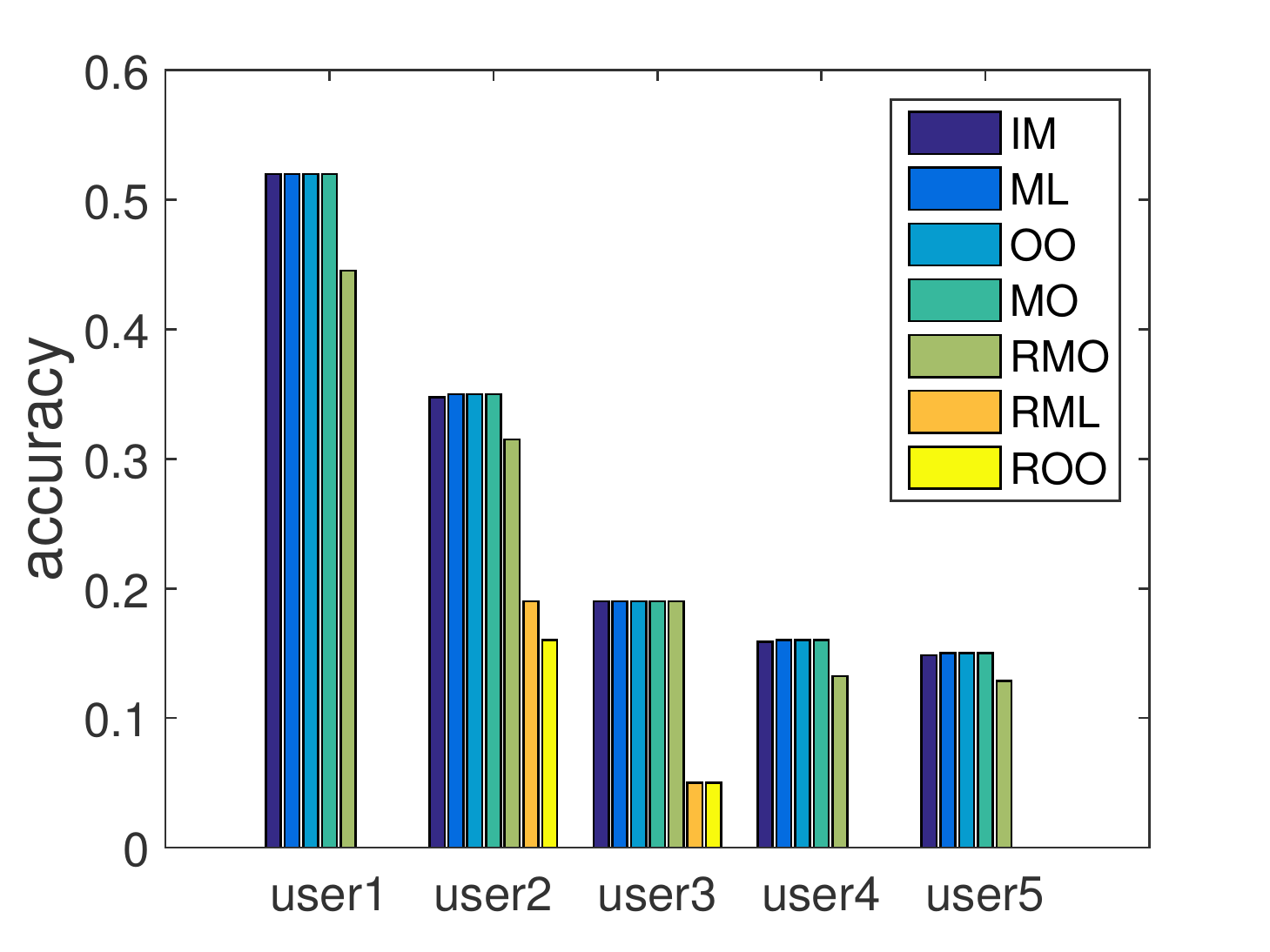}
\end{center}
 \vspace{-1.5em}
 \caption{\small Accuracy of advanced eavesdropper under $2$ chaffs. } \label{fig:accuracy_advanced_trace}
 \vspace{-.5em}
\end{figure}

We further evaluate the performance of an advanced eavesdropper under two chaffs. As shown in Fig.~\ref{fig:accuracy_advanced_trace}, the original strategies (IM, ML, OO, MO) are ineffective against this eavesdropper. In contrast, the robust strategies RML and ROO can substantially reduce the tracking accuracy.
Note that RMO performs poorly for reasons similar to those for MO under the basic eavesdropper.

\section{Conclusion and Discussions}\label{sec:Conclusion}

We studied the problem of protecting the location privacy of a mobile user using chaff services. Assuming that a cyber eavesdropper tracks the user by performing ML detection among observed service trajectories, we examined a range of chaff control strategies, from a baseline strategy to an optimal strategy. We proved that the optimal strategy and its online variation can reduce the tracking accuracy to zero when the entropy of the user's mobility is sufficiently high, while other straightforward strategies cannot. We further extended our strategies to improve their robustness against an advanced eavesdropper. Our evaluations highlighted the dependency of the eavesdropper's tracking accuracy on the user's mobility model, and verified the efficacy of our chaff-based defense mechanism in protecting the user's location privacy, even for users with highly predictable mobility.

\emph{Discussions:} We used chaff services as the defense mechanism to protect the user against untrusted MECs. Admittedly, running chaff services is expensive (see Section~\ref{subsec:Eavesdropper Model} for how we model the cost). While an ideal solution is to secure the entire system to eliminate the possibility of cyber eavesdropping, our solution provides additional protection when perfect security cannot be guaranteed (e.g., due to openness of MECs \cite{Satyanarayanan15CommMag,Schuster16}) or the consequence of successful eavesdropping is severe (e.g., in tactical applications).
While current solutions ignore the costs of running/migrating chaff services, our formulation can be extended to include constraints on such costs, and we leave a detailed study of the cost-privacy tradeoff to future work.

\tiny
\bibliographystyle{IEEEtran}
\bibliography{mybibSimplified}


\normalsize
\section*{Appendix}

\textbf{Proof of Lemma~\ref{lem:PUI vs PUM}}
\begin{proof}
Let $x^* := \argmax_{x\in\Lc} \pi(x)$. Then
\begin{align}
\sum_{x\in\Lc} \pi^2(x) - \max_{x\in\Lc}\pi(x) \hspace{-.25em} &=  \hspace{-.5em} \sum_{x\neq x^*}  \hspace{-.25em} \pi^2(x) - \pi(x^*)(1-\pi(x^*)) \nonumber\\
& \hspace{-.25em}=  \hspace{-.5em} \sum_{x\neq x^*} \pi(x)(\pi(x) - \pi(x^*)) \leq 0.
\end{align}
\end{proof}

\textbf{Proof of Lemma~\ref{lem:near-constant mean}}
\begin{proof}
First, we bound $\mbbE[c_{kw+i}|y_{(k-1)w+i}]$. Let $P^t(y|y')$ denote the $t$-step transition probability of $\{y_t\}_{t=1}^\infty$. By definition,
\begin{align}
\mbbE[c_{kw+i}|y_{(k-1)w+i}] &=\nonumber\\
 &\hspace{-6em} \sum_{y_{kw+i-1}\in \Lc^2} \hspace{-1em} P^{w-1}(y_{kw+i-1} | y_{(k-1)w+i}) g(y_{kw+i-1}),
\end{align}
and $\mbbE[c_{kw+i}]$ has a similar expression with $P^{w-1}(y_{kw+i-1} | y_{(k-1)w+i})$ replaced by $\pi(y_{kw+i-1})$. Thus,
\begin{align}
\Big| \mbbE[c_{kw+i}|y_{(k-1)w+i}] - \mbbE[c_{kw+i}] \Big| &\leq \nonumber\\
& \hspace{-10em} \sum_{y\in\Lc^2} |g(y)|\cdot \Big|
P^{w-1}(y | y_{(k-1)w+i}) - \pi(y) \Big|. \label{eq:proof |E[c|y]-E[c]|<=}
\end{align}
Since $w-1 = \tmix(\epsilon)$, we know by the definition of mixing time that $||P^{w-1}(\cdot | y_{(k-1)w+i}) - \pi ||_{\mbox{\tiny TV}} \leq \epsilon$ for all $y_{(k-1)w+i}$, where $||\cdot||_{\mbox{\tiny TV}}$ denotes the \emph{total variation distance} \cite{Levin09book}. This implies that $| P^{w-1}(y | y_{(k-1)w+i}) - \pi(y) | \leq \epsilon$ and $\sum_y | P^{w-1}(y | y_{(k-1)w+i}) - \pi(y) | \leq 2\epsilon$. Thus, the righthand side of (\ref{eq:proof |E[c|y]-E[c]|<=}) is upper-bounded by both $2\epsilon \max_y |g(y)|$ and $\epsilon \sum_y |g(y)|$, i.e., by $\epsilon \delta$.

Then, by the \emph{law of total expectation}, we have
\begin{align}
\mbbE[c_{kw+i}|c_{k'w+i}, \forall 0\leq k'<k] &= \nonumber\\
 & \hspace{-10em} \mbbE\Big[\mbbE[c_{kw+i}|y_{(k-1)w+i}] \Big| c_{k'w+i}, \forall 0\leq k'<k \Big],
\end{align}
which is bounded within $[\mbbE[c_{kw+i}] - \epsilon \delta,\: \mbbE[c_{kw+i}] + \epsilon \delta]$ based on the above result.
\end{proof}

\textbf{Proof of Lemma~\ref{lem:generalized Chernoff-Hoeffding}}
\begin{proof}
Define a new random variable $Y_t := X_t + \mu - \mbbE[X_t|X_1,\ldots,X_{t-1}]$. Then $\mbbE[Y_t|Y_1,\ldots,Y_{t-1}] = \mbbE[ \mbbE[Y_t | X_1,\ldots,X_{t-1}] | Y_1,\ldots,Y_{t-1} ] = \mu$ for all $t$. Further define $Z_t := (Y_t-a)/(b-a+\epsilon)$. Then $Z_t\in [0,\: 1]$ and $\mbbE[Z_t|Z_1,\ldots,Z_{t-1}]=(\mu-a)/(b-a+\epsilon) := \mu_z$.

Let $S^y_n$ denote the sum of $Y_1,\ldots,Y_n$ and $S^z_n$ denote the sum of $Z_1,\ldots,Z_n$. Then
\begin{align}
\Pr\{S_n\geq n(\mu+\Delta)\} &\leq \Pr\{S^y_n \geq n(\mu+\Delta)\} \label{eq:proof generalized Chernoff-Hoeffding 1}\\
&= \Pr\{ S^z_n \geq n\mu_z + {n\Delta \over b-a+\epsilon} \}  \nonumber \\
&\leq e^{-2n\Delta^2/(b-a+\epsilon)^2}, \label{eq:proof generalized Chernoff-Hoeffding 2}
\end{align}
where (\ref{eq:proof generalized Chernoff-Hoeffding 1}) is because $Y_t\geq X_t$, and (\ref{eq:proof generalized Chernoff-Hoeffding 2}) is by applying the {Chernoff-Hoeffding bound} \cite{Pollard84book} on $S^z_n$.
\end{proof}

\textbf{Proof of Theorem~\ref{thm:accuracy under CML}}
\begin{proof}
By definition of $c_t$, we have
\begin{align}
\PCML &\leq \Pr\{\sum_{t=1}^T c_t \geq 0\} \leq \Pr\{\sum_{t=2}^T c_t \geq -c_0\}\nonumber\\
&\leq \sum_{i=2}^{w+1} \Pr\{\sum_{k=0}^{T_i-1} c_{kw+i} \geq -{c_0\over w}\}, \label{eq:proof P_C(T)}
\end{align}
where the last step is by the union bound. Here $T_i := \lfloor (T-i)/w \rfloor+1 \geq T/w-1$.

By Lemma~\ref{lem:near-constant mean}, we know that $\mbbE[c_{kw+i}| c_{k'w+i}, \forall 0\leq k'<k] \in [-\mu-\epsilon \delta,\: -\mu+\epsilon \delta]$. Since $\mu-\epsilon \delta - c_0/(w T_i)\geq \mu-\epsilon \delta - c_0/(T-w) \geq 0$, we can apply Lemma~\ref{lem:generalized Chernoff-Hoeffding} to bound $\Pr\{\sum_{k=0}^{T_i-1} c_{kw+i} \geq -{c_0\over w}\}$ by
\begin{align}
& \Pr\{\sum_{k=0}^{T_i-1} c_{kw+i} \geq -T_i(\mu-\epsilon \delta) + T_i(\mu-\epsilon \delta - {c_0\over w T_i})\} \nonumber\\
&\leq \exp\left(-2T_i {(\mu-\epsilon \delta - {c_0\over w T_i})^2 \over (\cmax-\cmin+2\epsilon \delta)^2} \right) \nonumber\\
&\leq \exp\left(-2 \Big({T\over w}-1\Big) {(\mu-\epsilon \delta -{c_0\over T-w})^2 \over (\cmax-\cmin+2\epsilon \delta)^2}\right). \label{eq:proof sub-chain bound}
\end{align}
Plugging (\ref{eq:proof sub-chain bound}) into (\ref{eq:proof P_C(T)}) yields the final bound.
\end{proof}

\textbf{Proof of Theorem~\ref{thm:accuracy under MO}}
\begin{proof}
The proof is analogous to that of Theorem~\ref{thm:accuracy under CML}. First,
\begin{align}
\PMO(T) &\leq \Pr\{\sum_{t=1}^{T-1}c_t\geq -\cmax\} \leq \Pr\{\sum_{t=2}^{T-1}c_t\geq -c_0-\cmax\} \nonumber\\
&\leq \sum_{i=2}^{w'+1} \Pr\{\sum_{k=0}^{T_i-1}c_{kw'+i}\geq -{c_0+\cmax\over w'}\}, \label{eq:proof PMO(T)}
\end{align}
where $T_i:= \lfloor{T-i-1\over w'}\rfloor+1 \geq {T-w'-1\over w'}$.

By Lemma~\ref{lem:near-constant mean} (with $w$ replaced by $w'$ and $\delta$ replaced by $\delta'$), we have that $\mbbE[c_{kw'+i}|c_{k'w'+i},\forall 0\leq k'<k] \in [-\mu'-\epsilon\delta',\: -\mu'+\epsilon\delta']$. Since $\mu'-\epsilon\delta'-{c_0+\cmax\over T_i w'}\geq \mu'-\epsilon\delta'-{c_0+\cmax\over T-w'-1}\geq 0$, we can apply Lemma~\ref{lem:generalized Chernoff-Hoeffding} to obtain
\begin{align}
& \Pr\{\sum_{k=0}^{T_i-1}c_{kw'+i}\geq -{c_0+\cmax\over w'}\} \nonumber\\
&\leq \exp\left(-2T_i{(\mu'-\epsilon\delta'-{c_0+\cmax\over T_i w'})^2 \over (\cmax-\cmin+2\epsilon\delta')^2} \right) \nonumber\\
&\leq \exp\left(-2\Big({T-w'-1\over w'} \Big){(\mu'-\epsilon\delta'-{c_0+\cmax\over T-w'-1})^2\over (\cmax-\cmin+2\epsilon\delta')^2} \right). \label{eq:proof sub-chain bound 2}
\end{align}
Plugging (\ref{eq:proof sub-chain bound 2}) into (\ref{eq:proof PMO(T)}) yields the final bound.
\end{proof}

\textbf{Proof of Corollary~\ref{coro:time-average accuracy under MO}}
\begin{proof}
Note that Theorem~\ref{thm:accuracy under MO} holds for any value of $T$ and hence for $t=1,\ldots,T$. For $t\geq T_0$, the bound in Theorem~\ref{thm:accuracy under MO} applies, implying
\begin{align}
\PMO(t) &\leq w'\cdot \exp\left(-2\Big( {t-w'-1\over w'} \Big){(\mu'-\epsilon\delta'-{c_0+\cmax\over T_0-w'-1})^2\over (\cmax-\cmin+2\epsilon\delta')^2} \right) \nonumber\\
&= w' e^{-(t-w'-1)\alpha}.
\end{align}
For $t< T_0$, the bound does not apply, but we still have $\PMO(t)\leq 1$.

The overall tracking accuracy $\PMO:= {1\over T}\sum_{t=1}^T \PMO(t)$ thus satisfies
\begin{align}
\PMO &\leq {1\over T}\left(T_0-1 +\sum_{t=T_0}^T w'e^{(w'+1)\alpha}\cdot e^{-\alpha t} \right)\nonumber\\
&\leq {1\over T}\left(T_0-1 + {w' e^{\alpha(w'+1-T_0)} \over 1-e^{-\alpha}} \right).
\end{align}
\end{proof}

\end{document}